\documentclass[a4paper,fleqn,usenatbib]{mnras}

\usepackage{newtxtext,newtxmath}
\usepackage[T1]{fontenc}
\usepackage{ae,aecompl}
\usepackage{graphicx}
\usepackage{amsmath}
\usepackage{amssymb}

\title[Relativistic MHD KHI with magnetic polarization]{On the linear and non linear evolution of the Relativistic MHD Kelvin-Helmholtz instability in a magnetically polarized fluid}

\author[Oscar M. Pimentel and Fabio D. Lora-Clavijo]{
Oscar M. Pimentel$^{1}$\thanks{oscar.pimentel@correo.uis.edu.co (OMP)} and 
Fabio D. Lora-Clavijo$^{1}$\thanks{fadulora@uis.edu.co (FDLC)}
\\
$^{1}$Grupo de Investigaci\'on en Relatividad y Gravitaci\'on, Escuela de F\'isica, Universidad Industrial de Santander, A. A. 678,\\
Bucaramanga 680002, Colombia.\\
}

\date{Accepted XXX. Received YYY; in original form ZZZ}

\pubyear{2015}

\begin{document}
\label{firstpage}
\pagerange{\pageref{firstpage}--\pageref{lastpage}}
\maketitle

\begin{abstract}

The origin and strength of the magnetic field in some systems like active galactic nuclei or gamma-ray bursts is still an open question in astrophysics. A possible mechanism to explain the magnetic field amplification is the Kelvin-Helmholtz instability, since it is able to transform the kinetic energy in a shear flow into magnetic energy. Through the present work, we investigate the linear and non linear effects produced by the magnetic susceptibility in the development of the Kelvin-Helmholtz instability in a relativistic plasma. The system under study consists of a plane interface separating two uniform fluids that move with opposite velocities. The magnetic field in the system is parallel to the flows and the susceptibility is assumed to be homogeneous, constant in time, and equal in both fluids. In particular, we analyze the instability in three different cases, when the fluids are diamagnetic, paramagnetic, and when the susceptibility is zero. We compute the dispersion relation in the linear regime and found that the interface between diamagnetic fluids is more stable than between paramagnetic ones. We check the analytical results with numerical simulations, and explore the effect of the magnetic polarization in the non linear regime. We find that the magnetic field is more amplified in paramagnetic fluids than in diamagnetic ones. Surprisingly, the effect of the susceptibility in the amplification is stronger when the magnetization parameter is smaller. The results of our work make this instability a more efficient and effective amplification mechanism of seed magnetic fields when considering the susceptibility of matter.

\end{abstract}


\section{Introduction}

The Kelvin-Helmholtz instability (KHI) may occur when there is a sufficient relative velocity between two adjacent fluids. This shear in the velocity field is naturally generated in several scenarios, for instance, in the interaction between the solar wind plasma and the earth's magnetospheric plasma \citep{1984JGR....89..801M}, in the interaction between the relativistic outflows of the gamma-ray bursts (GRBs) and the surrounding medium, or in the relativistic jets coming from active galactic nuclei (AGN), where the shear is produced between the plasma, expelled form the disk-black hole system, and the cocoon \citep{2009ApJ...692L..40Z}. In these scenarios the KHI plays an important role as an energy dissipation and momentum transport mechanism. 

Now, when the shear flow has a magnetic field, the KHI may amplify the energy associated with that field \citep{1996ApJ...456..708M, 2014ApJ...793...60N} and produce magnetic reconnection events, which efficiently convert magnetic energy into kinetic, thermal and no-thermal energy \citep{2018MNRAS.476.4263T}. These magnetic effects give us an important insight to understand some open questions related to the strength of the magnetic fields in astrophysical systems. For instance, the KHI has been used by \cite{2009ApJ...692L..40Z} to explain the magnetic field strength required for the observed non-thermal GRBs emission. In another work, \cite{1981MNRAS.196.1051F} apply the KHI in a cylindrical geometry to describe the interaction between extragalactic jets and the interstellar medium. As it was pointed out by \cite{1979A&A....79..190F}, the KHI is a mechanism to accelerate relativistic electrons, and subsequently produce the observed synchrotron emission in relativistic jets.


In addition to the KHI, there are other powerful mechanisms to amplify magnetic fields, for instance, the dynamo effect which is capable to produce exponential growths of small seed magnetic fields and keep them for a long time. This mechanism has been succesfully used to explain the presence of long-live magnetic fields in different astrophysical bodies like planets, stars, galaxies, etc \citep{1978mfge.book.....M, 2018JPlPh..84d7304B}. Recently, the dynamo theory was applied by \cite{2014MNRAS.440L..41B} in resistive general relativistic MHD simulations to show the exponetinal growth of the magnetic field in a black hole accretion disc. In close relation with the dynamo process, the chiral magnetic effect has been considered as a promising mechanism for the generation and amplification of magnetic fields. It is a macroscopic quantum phenomena in which a chirality imbalance gives rise to an electric current in the same direction of the magnetic field \citep{2014PrPNP..75..133K}. This effect is possibly present in the quark-gluon plasma of proto-neutron stars. Recently, \cite{2018MNRAS.479..657D} include the chiral effect in the relativistic MHD (RMHD) equations to study the interplay between the dynamo, the chiral magnetic effect, and the magnetic reconnection. Another well-known mechanism is the magnetorotational instability \citep{1991ApJ...376..214B} which is produced through the interaction of a differentially rotating fluid and a magnetic field. This magnetohydrodynamic (MHD) instability grows rapidly and therefore, it could contribute with the dynamo amplification of very weak magnetic fields \citep{2002ApJ...577..524Q}.



Now, the free currents in the plasmas are not the only sources of the observed magnetic fields in nature, there is another contribution associated with the dipolar structure of matter, i.e. with the electrons orbiting around the nucleus and spinning around themselves. When there is a net alignment of magnetic dipoles, the fluid is said to be magnetically polarized. This polarization interacts with the magnetic field, and generates an additional field \citep{citeulike:4033945}.

The magnetic polarization of matter has already been considered to study the structure of neutrons stars. In particular, \cite{1982JPhC...15.6233B} explored the effect of the free electrons magnetic susceptibility on the surface properties of the star crust. In this paper the authors show that the magnetization do not contribute significantly with the equillibrium structure but it could be related to observable effects. Later, \cite{2010ApJ...717..843S} investigated the possibility that the Soft Gamma-Ray Repeaters and the Anomalous X-ray Pulsars were observational evidence for the magnetic domain formation in neutron stars. They conclude that magnetic domains can be formed in the crust and core of magnetars. Moreover, as it was mentioned by \cite{2016PASP..128j4201W}, neutron stars are ideal astrophysical laboratories to test the de Haas-van Alphen effect \citep{de1930dependence}, in which the magnetic susceptibility oscilates as the magnetic field is increased, and also to explain the diamagnetic phase transition, which is associated with magnetic domain formation. On the other hand, \cite{2015PhRvD..92l4034K} found a magnetic field amplification of $\approx 10^3$ times the initial magnetic field due to the KHI in a binary neutron star merger. This result is even more important nowadays due to the signal GW170817, recognized as the first gravitational wave detection from the merger of two neutron stars \citep{2017PhRvL.119p1101A}. Then, since the dipolar structure of matter and the KHI could be astrophysically relevant, we will explore in this paper the role of the magnetic polarization of matter in the linear and non linear phases of the KHI.

The linear stage of the MHD KHI in the plane-parallel geometry is well-analyzed in the incompresible case by \cite{1961hhs..book.....C} and in the compressible case by \cite{1983JGR....88..841P}. The relativistic case was initially addressed by \cite{1980MNRAS.193..469F} in the limit where the Alfv$\grave{\text{e}}$n velocity is much smaller than the speed of light. More recently, the MHD KHI was generalized in \cite{2008A&A...490..493O} by allowing any Alfv$\grave{\text{e}}$n velocity. In this last paper the authors compute the analytical dispersion relation in the frames where the fluids are at rest, and found that the modes, propagating in the same direction of the flow velocity, destabilize the interface only at low magnetizations. On the other hand, the non linear evolution of the instability has been studied in the non-relativistic regime by \cite{1984JGR....89..801M}. In this work, the authors analyze the cases where the magnetic field is parallel and perpendicular to the flow velocity, and found that in the parallel case the magnetic energy can be amplified by vortical motions that twist and compress the field. Later, \cite{1996ApJ...456..708M} extended the previous work in the parameter space and in the simulation time, and found that the field amplification is actually a transient phenomenon which eventually decay in a long enough time. This result makes sense since the dynamo process can not operate in a two-dimensions space \citep{1978mfge.book.....M}, and it is necessary a 3D simulation to explore the dynamo effect triggered by the KHI \citep{2009ApJ...692L..40Z}. The non linear stages of the relativistic RMHD KHI were first obtained by \cite{2006A&A...454..393B} in order to compare the growth rates and the synchrotron emission of the instability with the observed variable features of the Crab Nebula.

Recently, in \cite{2018ApJ...861..115P} we presented the eigenvalue structure of the general RMHD with magnetic polarization, and showed that the magnetic susceptibility modifies the Alfv$\grave{\text{e}}$n and the magneto-acoustic modes. With this characteristic structure, it is possible to follow \cite{2008A&A...490..493O} and make a linear analysis of the instability in a magnetically polarized fluid. This analysis allows us to determine the effect of the magnetic susceptibility in the growth rate of disturbances. On the other hand, in \cite{2018ApJ...861..115P} we present the RMHD equations with magnetic polarization in the 3+1 decomposition, and implement them in the CAFE code \citep{2015ApJS..218...24L} with the objective of studying the diamagnetic and paramagnetic properties of matter in astrophysical scenarios. For instance, in black hole accretion discs, for which we obtain in \cite{2018arXiv180807400P} a new analytical solution of magnetically polarized tori that can be useful as initial configuration to obtain the accretion process of diamagnetic and paramagnetic fluids. Therefore, it is also possible to simulate the linear and non linear evolution of the RMHD KHI. These simulations will be very useful to check the results of the linear theory and explore the effect of the magnetic susceptibility in the magnetic field amplification and saturation state which are non linear features of the instability.

The organization of the paper is the following. In section \ref{sec2} we introduce the energy-momentum tensor, the RMHD equations in conservative form, and the initial configuration of the magnetically polarized fluid. In section \ref{sec3} we present the linear analysis of the KHI, so we compute the equations that describe the evolution of small perturbations and find the dispersion relation in the frames where the fluids are at rest. Direct solutions to this equation give us the growth rates of the instability in terms of physical parameters. In section \ref{sec4} we present the first simulations of the linear and non-linear phases of the RMHD KHI with magnetic polarization. Finally, in section \ref{sec5} we summarize and discuss the main results of this work.

Throughout the paper, the greek indices represent the spacetime coordinates while the latin indices denote the space coordinates. In particular, $\mu,\nu=(t,x,y,z)$, where $t$ is the temporal coordinate and $i,j,k=(x,y,z)$ are the usual cartesian coordinates.


\section{Basic equations and Initial Setup}
\label{sec2}

In order to study the linear and non linear dynamics of the RMHD KH instability in the presence of a magnetically polarized fluid, we introduce the conservation laws 
\begin{equation}
\partial_{\mu}T^{\mu\nu}=0, \hspace{5mm} \partial_{\mu}(\rho u^{\mu})=0,
\label{conservation_laws}
\end{equation}
and the relevant Maxwell equations given by,
\begin{equation}
\partial_{\mu}(u^{\mu}b^{\nu}-b^{\mu}u^{\nu})=0,
\label{relevant_maxwell}
\end{equation}
where $\rho$ and $b^{\mu}$ are the rest mass density and the magnetic field, both measured by a comoving observer whose four-velocity vector is $u^{\mu}$. The energy-momentum tensor, $T^{\mu\nu}$, defined in terms of the Faraday and magnetization tensors, $F^{\mu\nu}$ and $M^{\mu\nu}$ respectively, takes the form \citep{Maugin:1978tu, 2010PhRvD..81d5015H, 2015MNRAS.447.3785C}
\begin{align}
T^{\mu\nu}=&\rho hu^{\mu}u^{\nu}+p\eta^{\mu\nu}-\frac{1}{\mu_{0}}\left(F^{\mu\alpha}F_{\alpha}^{\ \nu}+\frac{1}{4}\eta^{\mu\nu}F_{\alpha\beta}F^{\alpha\beta}\right) \nonumber \\& + \frac{1}{2}\left(F^{\nu}_{\ \alpha}M^{\alpha\mu}+F^{\mu}_{\ \alpha}M^{\alpha\nu}\right),.
\label{energy_tensor_prev}
\end{align}
where $h$ is the specific enthalpy, $p$ is the thermodynamic pressure, and $\eta^{\mu\nu}=\text{diag}(-1,1,1,1)$ are the components of the metric tensor. In the ideal MHD approximation $F^{\mu\nu}$ and $M^{\mu\nu}$ can be decomposed as
\begin{equation}
F^{\mu\nu}=\epsilon^{\mu\nu\alpha\beta}b_{\alpha}u_{\beta}, \hspace{5mm} M^{\mu\nu}=\epsilon^{\mu\nu\alpha\beta}m_{\alpha}u_{\beta},
\label{decomposed}
\end{equation}
being $\epsilon^{\mu\nu\alpha\beta}$ the Levi-Civita tensorial density and $m^{\mu}$ the magnetization vector.

Throughout this paper we will assume that the magnetic field and the magnetization vector satisfy the linear constitutive relation \citep{griffiths2005introduction},
\begin{equation}
m^{\mu}=\frac{\chi}{\mu_0}b^{\mu},
\label{constitutive}
\end{equation}
where $\chi=\chi_m/(1+\chi_m)$ and $\chi_m$ is the magnetic susceptibility. When this parameter is negative ($\chi_m<0$) the fluid is diamagnetic and when it is positive ($\chi_m>0$) the fluid is paramagnetic. Therefore, the effects of the magnetic polarization will be introduced through $\chi_m$. Replacing (\ref{decomposed}) and (\ref{constitutive}) in (\ref{energy_tensor_prev}) we obtain,
\begin{equation}
T^{\mu\nu}=\rho h^{*}u^{\mu}u^{\nu}+p^{*}\eta^{\mu\nu}-\frac{1}{\mu_{0}}(1-\chi)b^{\mu}b^{\nu},
\label{energy_tensor}
\end{equation} 
with
\begin{equation}
\rho h^{*}=\rho h+(1-\chi)b^{2}/\mu_{0}, \hspace{5mm} p^{*}=p+(1-2\chi)(b^{2}/2\mu_{0}),
\label{h_rho}
\end{equation}
and $b^{2}=b^{\mu}b_{\mu}$. It is worth mentioning that the net effect of $\chi_m$ can't be understood as a renormalization of the magnetic field, $b^{\mu}\rightarrow \sqrt{1-\chi}b^{\mu}$. A discussion about the differences between the usual RMHD but with a renormalized magnetic field and the RMHD with the complete form of the energy-momentum tensor given in (\ref{energy_tensor}) is presented in the appendix \ref{apenA}.

The system of equations (\ref{conservation_laws}-\ref{relevant_maxwell}) can be transformed into the conservative system of evolution equations,
\begin{equation}
\partial_{t}\vec{U}+\partial_{i}\vec{F}^{i}=0, \label{conservative} 
\end{equation}
and the divergence-free condition $\partial_{i}B^{i}=0$, being $B^{i}$ the laboratory frame magnetic field. In the system (\ref{conservative}), $\vec{U}$ is the conservative variables vector,
\begin{equation}
\vec{U}=\left[
\begin{array}{c}
\vspace{2mm}
D\\
\vspace{2mm}
S_{j}\\
\vspace{2mm}
\tau\\
B^{k}
\end{array}
\right]=\left[
\begin{array}{c}
\vspace{2mm}
\rho \Gamma\\
\vspace{2mm}
\rho h^{*}\Gamma^{2}v_{j}-\frac{1}{\mu_{0}}(1-\chi)b^{t}b_{j}\\
\vspace{2mm}
\rho h^{*}\Gamma^{2}-p^{*}-\frac{1}{\mu_{0}}(1-\chi)(b^{t})^{2}-\rho \Gamma\\
B^{k}
\end{array}
\right],
\label{U}
\end{equation}
and $\vec{F}^{i}$ is the flux vector,
\begin{equation}
\vec{F}^{i}=\left[
\begin{array}{c}
\vspace{2mm}
D v^{i}\\
\vspace{2mm}
S_{j}v^{i}+p^{*}\delta_{j}^{i}-\frac{b_{j}B^{i}}{\mu_{0}\Gamma}(1-\chi)\\
\vspace{2mm}
\tau v^{i}+p^{*}v^{i}-\frac{b^{t}B^{i}}{\mu_{0}\Gamma}(1-\chi)\\
v^{i}B^{k}-v^{k}B^{i}
\end{array}
\right],
\label{F}
\end{equation}
where $\Gamma$ is the Lorentz factor between the laboratory and comoving frames, and $v^{j}$ is the tree-velocity in the laboratory frame. The four-velocity $u^{\mu}$ and the three-velocity are related through the equation $u^{\mu}=(\Gamma,\Gamma v^{i})$. In a similar way, the magnetic field in the comoving frame, $b^{\mu}$ can be written in terms of $B^{i}$ as
\begin{equation}
b^{\mu}=\left[\Gamma B^{i}v_{i},\frac{B^{j}}{\Gamma}+\Gamma(B^{i}v_{i})v^{j}\right].
\label{bBrelation}
\end{equation}
A more complete description of the conservative form of the RMHD equations with magnetic polarization can be found in \cite{2018ApJ...861..115P}.

The initial configuration we have used to find out the role of $\chi_m$ in the RMHD KH instability consists of a magnetically polarized shear flow in the $x-y$ plane which moves along the $x-$direction. In the region $y>0$ the fluid moves with a velocity $\vec{v}=V_0\hat{\imath}$, and in the region $y<0$ with $\vec{v}=-V_0\hat{\imath}$, so the interface between the fluids is located in $y=0$. Additionally, we suppose that the flow has an uniform density and pressure profiles, and that it is affected by a constant magnetic field of the form $\vec{B}=B_{x}\hat{\imath}$. We have adopted the parameters used by \cite{2006A&A...454..393B} because they are motivated in the local conditions observed in pulsar wind nebulae, so they are of astrophysical relevance. In particular, we assume that $V_0=0.25$, $p/\rho=20$ (with $\rho=1$) and $\gamma=4/3$, which are appropriate values for a hot relativistic plasma. We also introduce the magnetization parameter $\beta_{x}=B_{x}^{2}/2p$ as the ratio of magnetic to thermodynamic pressures. Finally, we assume that the magnetic susceptibility, $\chi_m$ is constant in all the space and during all the evolution.

The configuration described in the last paragraph is unstable under certain conditions of density, pressure, magnetic field, etc. In nature, random motions in the fluids excite unstable modes that grow in time and end up in a turbulent state. The growth rate of those modes can be computed analytically in the linear phase, so it is possible to determine the effect of $\chi_m$ in the early stages of the KHI. This is precisely the main objective of the next section.


\section{Linear analysis}
\label{sec3}

The linear evolution of the relativistic KHI in a magnetically polarized fluid is obtained by closely following the work of \cite{2008A&A...490..493O}. In this paper the authors work in the comoving reference frames of the fluids because in that way the dispersion relation takes a very simple form. The system, consisting of two fluids moving in opposite directions, is described by the variables $\tilde{\Psi}_{\pm}=(\tilde{\rho}_{_\pm}, \tilde{p}_{_\pm}, \tilde{v}^i_{_\pm}, \tilde{B}^{i}_{_\pm})$, where the tilde is used for the quantities measured in the rest frames of the fluids (comoving frames), and the $\pm$ sign represents the upper ($y>0$) and lower ($y<0$) fluids respectively. We add a small perturbation, $\delta\tilde{\Psi}_{\pm}$, to the initial configuration so the new state can be written as
\begin{equation}
\tilde{\Psi}_{\pm}=\tilde{\Psi}_{0_{\pm}}+\delta\tilde{\Psi}_{\pm},
\label{perturbed}
\end{equation}
being $\tilde{\Psi}_{0_{\pm}}$ the physical variables in the unperturbed state. The space-time dependence of the disturbance is assumed as
\begin{equation}
\delta\tilde{\Psi}_{\pm}\propto\exp{\left[i(\tilde{k}_{\pm}\tilde{x}+\tilde{l}_{\pm}\tilde{y}+
\tilde{m}_{\pm}\tilde{z}-\tilde{\omega}_{\pm}\tilde{t})\right]},
\label{perturbation_ecu}
\end{equation}
where $\tilde{k}_{\pm}$, $\tilde{l}_{\pm}$, $\tilde{m}_{\pm}$ are the wave numbers in each spatial direction and $\tilde{\omega}_{\pm}$ is the frequency. According to the last equation, the unstable modes arise when the frequency is a complex number with a positive imaginary part. The wave numbers and the frequency in the comoving frames are related to the same quantities in the laboratory frame, $k, l$, and $m$, through the Lorentz transformations, 
\begin{equation}
\tilde{\omega}_{\pm}=\Gamma(\omega\mp kV_0),
\label{omegaLortz}
\end{equation}  
\begin{equation}
\tilde{k}_{\pm}=\Gamma(k\mp\omega V_0),
\label{kLortz}
\end{equation}
\begin{equation}
\tilde{l}_{\pm}=l_{\pm}, \hspace{3mm} \tilde{m}_{\pm}=m,
\label{lmLortz}
\end{equation}
where $\Gamma=1/\sqrt{1-v^2}$.

The dispersion relation can be obtained through the characteristic equation for the RMHD with magnetic polarization \citep{2018ApJ...861..115P},
\begin{equation}
(\rho h)^{-1}a^{2}A^{2}N_{4}Q=0,
\label{characteristic_equation}
\end{equation}
where
\begin{align}
&A=[\rho h+(1-\chi)b^{2}]a^{2}-(1-\chi)B^{2},\\
&N_{4}=f_{1}a^{4}-f_{2}a^{2}G-f_{3}a^{2}B^{2}+f_{4}B^{2}G,\label{n4}\\
&Q=b^{4}\chi^{2}-b^{2}(b^{2}+2\rho h)\chi+\rho h(\rho h+b^{2}),
\end{align}
and $a=u^{\mu}\phi_{\mu}$, $B=b^{\mu}\phi_{\mu}$, $G=\phi_{\mu}\phi^{\mu}$. The term $N_4$ is defined by the functions
\begin{align}
&f_1=(e'_{p}-1)\rho h+(e'_{p}+1)\chi b^{2},\\
&f_2=\rho h+(1-2\chi)b^{2}e'_{p}-\chi b^{2},\\
&f_3=(e'_{p}+1)\chi,\\
&f_4=1-\chi.
\end{align}
The characteristic equation describes the propagation speeds of a disturbance with a wave vector $\phi_\mu=(-\omega,k,l,m)$. These speeds are computed by equaling to zero only the functions $a$, $A$, and $N_4$, because $Q$ does not depend on the wave vector. 

By writing (\ref{characteristic_equation}) in the comoving frame where $u^0=1$, $u^{i}=0$, $b^{0}=0$, and $b^{i}=B^{i}$ we find that $a=-\tilde{\omega}$, $B=B_{x_0}\tilde{k}$, and $G=-\tilde{\omega}^2+\tilde{k}^2+\tilde{l}^2+\tilde{m}^2$. In this way, the solutions to $a=0$, which correspond to the entropic modes, are always real and do not contribute to the instability. The Alfv$\grave{\text{e}}$n modes, obtained from $A=0$, are solutions to the equation $\tilde{\omega}_{\pm}^2-V_{A}^{2}\tilde{k}_{\pm}^2=0$, where
\begin{equation}
V_{A}^{2}=\frac{(1-\chi)B_{x_{0}}^2}{\rho_0 h_0+(1-\chi)B_{x_{0}}^2}
\label{alfven_velocity_kh}
\end{equation}
is the Alfv$\grave{\text{e}}$n velocity, which is modified by the magnetic susceptibility of the fluid. Since the frequencies associated with the Alfv$\grave{\text{e}}$n modes are always real, they do not contribute to the instability. Conversely, the magneto-acoustic modes ($N_4=0$), whose frequencies are solutions to the equation,
\begin{equation}
\frac{\tilde{\omega}_{\pm}^{4}}{(\tilde{k}_{\pm}^{2}+\tilde{l}^{2}_{\pm}+\tilde{m}^{2}_{\pm})^{2}}-\tilde{\mu}_{\pm}
\frac{\tilde{\omega}_{\pm}^{2}}{\tilde{k}_{\pm}^{2}+\tilde{l}^{2}_{\pm}+\tilde{m}^{2}_{\pm}}+
\tilde{\nu}_{\pm}^{2}=0,
\label{magnetoacustic_equation}
\end{equation}
can be imaginary, so they correspond to unstable states. In the last equation,
\begin{equation}
\tilde{\mu}_{\pm}=C_s^2+V_A^2-V_A^2\left[C_s^2+(1+C_s^2)\frac{\chi}{1-\chi}\right]\frac{\tilde{l}_{\pm}^2+\tilde{m}_{\pm}^2}{\tilde{k}_{\pm}^{2}+\tilde{l}^{2}_{\pm}+\tilde{m}^{2}_{\pm}},
\label{acus_1}
\end{equation}  
and
\begin{equation}
\tilde{\nu}_{\pm}^2=V_a^2C_s^2\frac{\tilde{k}_{\pm}^{2}}{\tilde{k}_{\pm}^{2}+\tilde{l}^{2}_{\pm}+\tilde{m}^{2}_{\pm}},
\label{acus_2}
\end{equation}
being $C_s=\sqrt{\gamma p_0/\rho_0h_0}$ the sound speed.

Now, the linearized momentum equations in the $y$ and $z$ directions, and the induction equation can be obtained in terms of $\chi_m$ by replacing (\ref{perturbation_ecu}) in (\ref{conservative}-\ref{F}). The linearization process leads to the following system: 
\begin{align}
&(1-\chi)B_{x_{0}}\frac{\tilde{k}_{\mp}}{\Gamma_0}\delta B_{y_{\pm}}+\left[(1-\chi)\omega B_{x_{0}}^2+\rho_0 h_0 \Gamma_0^2\tilde{\omega}_{\mp}\right]\delta v_{y_{\pm}}-\nonumber\\
&-\left[\delta p_{\pm}+(1-2\chi)B_{x_{0}}\delta B_{x_{\pm}}\right]m=0, \label{eq1}\\
&(1-\chi)B_{x_{0}}\frac{\tilde{k}_{\mp}}{\Gamma_0}\delta B_{z_{\pm}}+\left[(1-\chi)\omega B_{x_{0}}^2+\rho_0 h_0 \Gamma_0^2\tilde{\omega}_{\mp}\right]\delta v_{z_{\pm}}-\nonumber\\
&-\left[\delta p_{\pm}+(1-2\chi)B_{x_{0}}\delta B_{x_{\pm}}\right]m=0, \label{eq2}\\
&(\omega\mp kV_0)\delta B_{x_{\pm}}-B_{x_{0}}(l_{\pm}\delta v_{y_{\pm}}+m\delta v_{z_{\pm}})=0, \label{eq3}\\
&(\omega\mp kV_0)\delta B_{y_{\pm}}+kB_{x_{0}}\delta v_{y_{\pm}}=0, \label{eq4}\\
&(\omega\mp kV_0)\delta B_{z_{\pm}}+kB_{x_{0}}\delta v_{z_{\pm}}=0. \label{eq5}
\end{align}
Equations (\ref{eq3}-\ref{eq5}) are useful to write $\delta B_{i_{\pm}}$ in terms of the velocity perturbations, so we can express the total pressure perturbation as
\begin{equation}
\delta p_{t_{\pm}}=\delta p_{\pm}+(1-2\chi)\frac{l_{\pm}B_{x_0}^2\delta v_{y_{\pm}}}{\omega\mp kV_0}.
\label{pres_pert}
\end{equation}
Then, using (\ref{eq1}) and the matching conditions at the interface,
\begin{equation}
p_{t_{+}}=p_{t_{-}}$, \hspace{3mm} $\frac{\delta v_{y_{+}}}{\omega-kV_0}=\frac{\delta v_{y_{-}}}{\omega+kV_0},
\label{matchin}
\end{equation}
we obtain,
\begin{equation}
\frac{l_+}{l_-}=\frac{\rho_0h_0\Gamma_0^2(\omega-kV_0)^2+(\omega^2-k^2)(1-\chi)B_{x_0}^2}{\rho_0h_0\Gamma_0^2(\omega+kV_0)^2+(\omega^2-k^2)(1-\chi)B_{x_0}^2}.
\label{l_conditions}
\end{equation}
With this result and the expression for $l_{\pm}^2$ that is calculated from the dispersion relation (\ref{magnetoacustic_equation}), we can compute numerically the imaginary part of the frequency, Im($\omega$), which determine the growth rate of the instability.

Even though the procedure also admits perturbations in an arbitrary direction of the interface, the analysis will be restricted to the case $m=0$. It means that we will only study perturbations propagating parallel to the $x-y$ plane. Moreover, following \cite{2006A&A...454..393B}, we will consider in this work the instability of the mode with the longest possible wavelength, corresponding to $k=1$. Then, in the entire domain there will be only one wavelength. We make this choice because we are mainly interested in the large-scale features of the instability.

\begin{figure}
\begin{tabular}{r}
\includegraphics[height=2.65in]{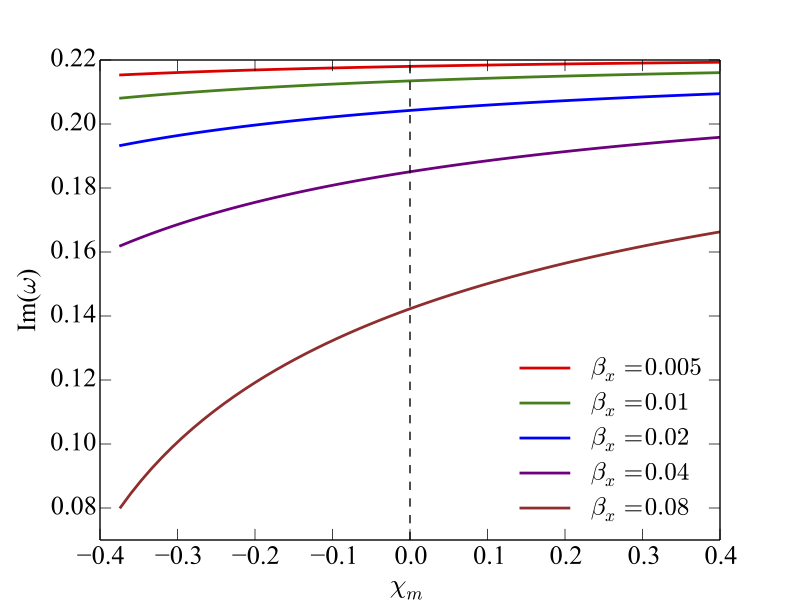}
\end{tabular}
\caption{Growth rate Im($\omega$) as a function of the magnetic susceptibility $\chi_m$ for different values of $\beta_x$.}
\label{growth_chi}
\end{figure}

In figure (\ref{growth_chi}) we plot the growth rate, Im($\omega$), as a function of the magnetic susceptibility of the fluids for different values of the magnetization parameter. We notice that the diamagnetism ($\chi_m<0$) tends to stabilize the interface between the fluids, but when the paramagnetism is present ($\chi_m>0$), the instability grows faster than the original case with $\chi_m=0$, represented by the vertical dashed line. This behavior becomes increasingly important with the magnetization parameter, so we expect to observe the linear effect of $\chi_m$ in the cases with $\beta_x=0.02$, $\beta_x=0.04$, and $\beta_x=0.08$. Conversely, we shouldn't see any differences in the approximately hydrodynamic cases where $\beta_x=0.005$ and $\beta_x=0.01$. 

\begin{figure}
\begin{tabular}{c}
\includegraphics[height=2.65in]{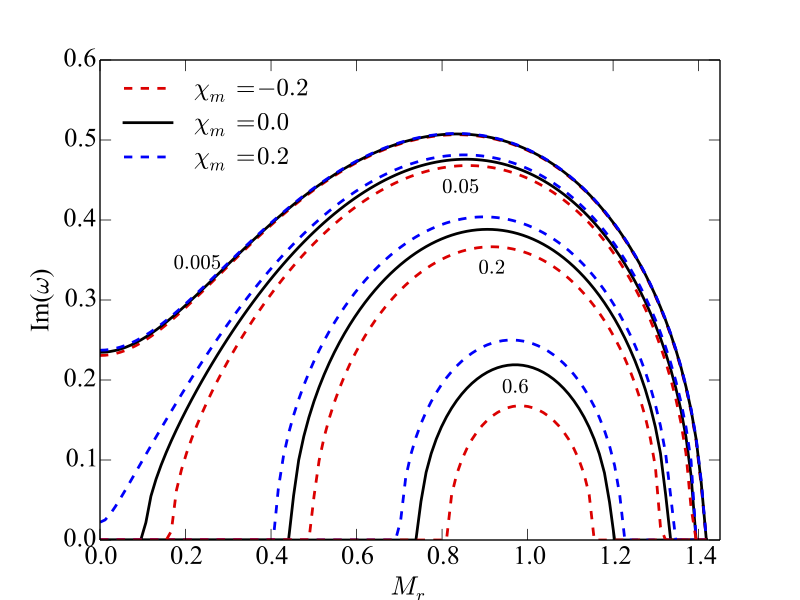}
\end{tabular}
\caption{Growth rate Im($\omega$) as a function of the relativistic Mach number $M_r$ for different values of $\beta_x$.}
\label{growth_mr}
\end{figure}

Sometimes it is useful to analyze the stability properties of the system in terms of the relativistic Mach number, 
\begin{equation}
M_r=\frac{V_0/\sqrt{1-V_0^{2}}}{C_s/\sqrt{1-C_s^2}}.
\label{mach}
\end{equation}
In the usual MHD case, where the fluids are not polarized, there is an interval of $M_r$ for which the system is unstable \citep{2008A&A...490..493O}. The width of this interval is reduced when we consider a higher magnetic field because it acts as a surface tension stabilizing the interface. In figure (\ref{growth_mr}) we show the effect of the magnetic susceptibility on the interval of $M_r$ with a non-zero growth rate. The black curves correspond to the usual MHD case with $\chi_m=0$, the blue-dashed lines are for paramagnetic fluids, and the red-dashed ones are for the diamagnetic cases. Next to each set of curves we specify the corresponding magnetization parameter value. When $\beta_x=0.005$, the magnetic polarization does not produce a significant change in the growth rates, but in the other cases ($\beta_x=0.05, 0.02, 0.06$), the magnetic susceptibility changes significantly the values of Im($\omega$), specially when $M_r$ approaches to its first unstable value. In particular, the paramagnetism in matter tends to destabilize the system by increasing the interval of $M_r$ for which the Kelvin-Helmholtz instability appears. Conversely, the diamagnetism reduce the values of $M_r$ with a non-zero growth rate. Now, all substances in nature are diamagnetic, but when the paramagnetism is present it is usually dominant. Then, any magnetized shear flow in the configuration we are studying in this work should be more stable than the well-known case without magnetic polarization, unless the paramagnetism is present.

In the next section we will use numerical simulations to test our analytical results and to obtain the non-linear evolution of the Kelvin-Helmholtz instability as a function of the magnetic susceptibility. The non-linear regime is important because it is associated with the magnetic field amplification, the saturation state, and the turbulent behavior.

\section{Non linear evolution}
\label{sec4}

To numerically study the non linear evolution of the RKHI in a magnetically polarized fluid, we assume a velocity profile that varies on the $y-$direction according to the equation,
\begin{equation}
v_{x}=-V_{0}\tanh(y/\alpha),
\label{vx_profile}
\end{equation}
where $\alpha$ is a constant associated with the thickness of the transition layer between the lower ($y<0$) and upper ($y>0$) fluids. As it is mentioned in \cite{1996ApJ...456..708M}, this kind of profile stabilizes perturbations with wavelengths comparable with the numerical resolution and appropriately reproduces the initial configuration we use in the linear analysis. Then, we can compare the numerical results with those of the previous section. Now, to trigger the instability, we introduce the following disturbance to the initial flow velocity,
\begin{equation}
v_{y}=v_{y_{0}}\sin{(kx)}\exp{[-(y/\sigma)^{2}]},
\label{disturbance}
\end{equation}  
where $v_{y_{0}}$ is the amplitude of the perturbation and $\sigma$ is a parameter that describes the width of the perturbation in the $y-$direction. The perturbation's amplitude is such that $v_{y_{0}}\ll V_{0}$.

The perturbed initial configuration evolves according to the non-linear and time-dependent equations (\ref{conservative}-\ref{F}). This system of equations is numerically solved by means of the CAFE code \citep{2015ApJS..218...24L}, which was created to study the evolution of a relativistic test fluid threaded by a magnetic field in the vicinity of a compact object. This code has been successfully used to simulate accretion processes onto black holes \citep{2015ApJS..219...30L, 2016MNRAS.460.3193C, 2017MNRAS.471.3127C}. Recently, CAFE was modified in \cite{2018ApJ...861..115P} with the aim of analysing the effect of the fluid's magnetic polarization in some astrophysical scenarios.

The numerical methods implemented in CAFE are described in detail in \cite{2015ApJS..218...24L}, and the modifications that include the magnetic polarization of the matter are fully presented in \cite{2018ApJ...861..115P}. In summary, CAFE uses the HLLE Riemann solver along with a spatial reconstructor to compute the numerical fluxes (\ref{F}) on both sides of the grid points (intercells). In this work we use the MC reconstructor because it adequately reproduces the analytical growth rates of the usual RMHD KH instability (the cases with $\chi_m=0$). Now, since the HLLE method only requires the eigenvalue structure of the system of equations, we computed these quantities by following the Anile procedure and included them in the new version of CAFE with magnetic polarization. Finally, it is important to mention that CAFE satisfies the divergence-free condition for the magnetic field by means of the flux constraint transport method \citep{1988ApJ...332..659E}.

The equations presented in (\ref{conservative}-\ref{F}) are solved in a 2D numerical domain of size $D=2\pi /k$ in the $x-$direction and $2D$ in the $y-$direction. In addition, we define the resolution of our simulations by taking $N_{x}=180$ and $N_{y}=360$, being $N_x$ and $N_y$ the number of grid points along each spatial direction. Now, we set the initial velocity profile (\ref{vx_profile}) and the perturbation (\ref{disturbance}) with the parameters $\alpha=D/100$, $\beta=D/10$ and $v_{y_{0}}=0.01V_0$. We also rescale the spatial distances and the time with the values $D$ and $D/c$, respectively, in such a way that the spatial domain becomes $[0,1]\times [-1,1]$. Finally, as in \cite{1996ApJ...456..708M}, we use periodic boundary conditions at $x=0$ and $x=1$, and free outflow conditions at $y=-1$ and $y=1$.

\begin{figure*}
\begin{tabular}{ccc}
\includegraphics[height=2.9in]{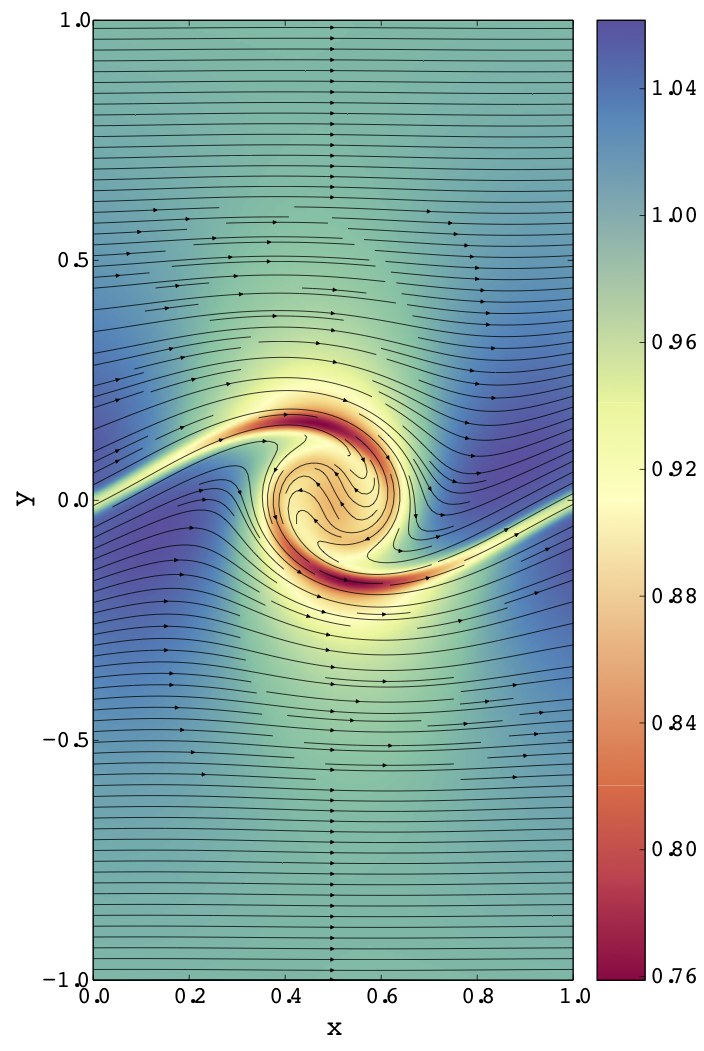} & 
\includegraphics[height=2.9in]{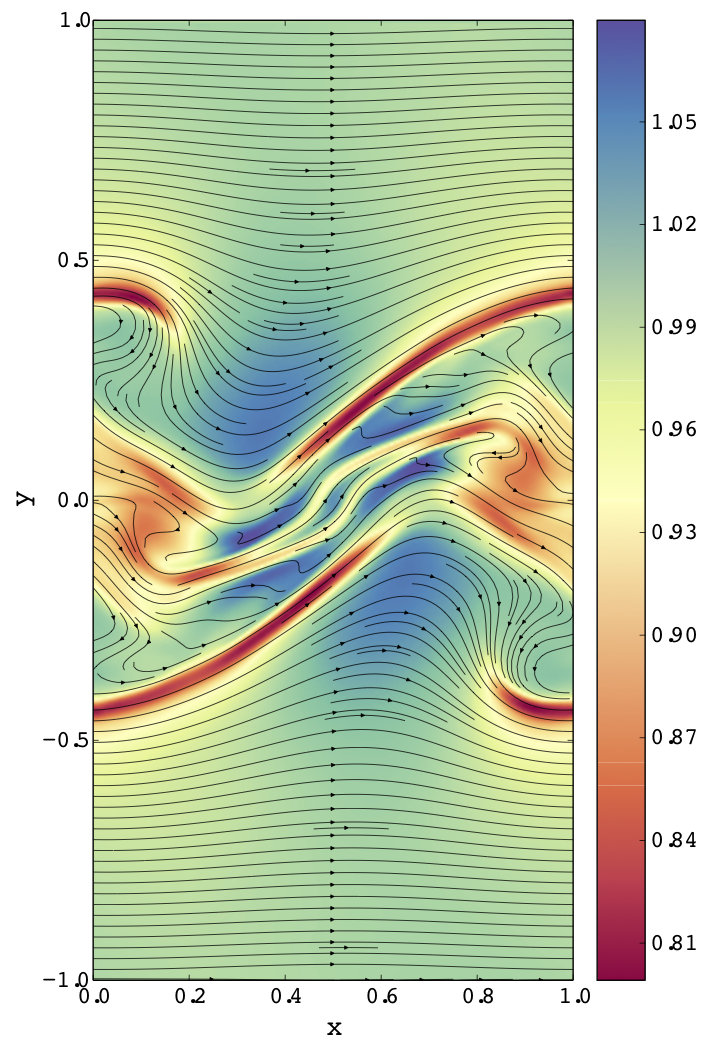} &
\includegraphics[height=2.9in]{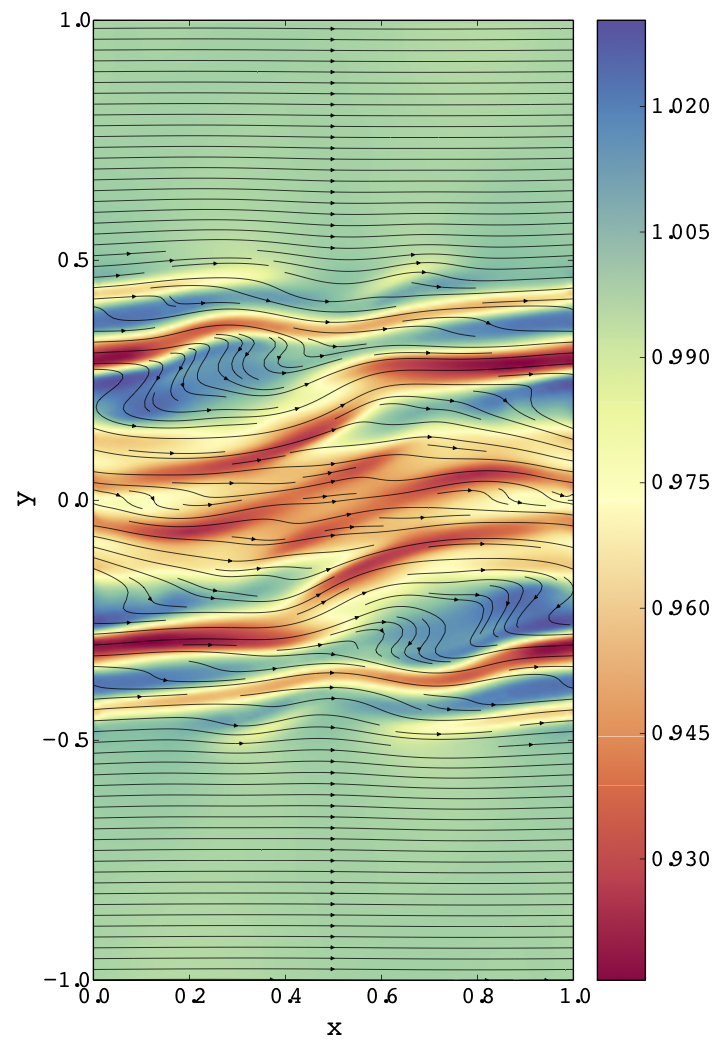}\\
\includegraphics[height=2.9in]{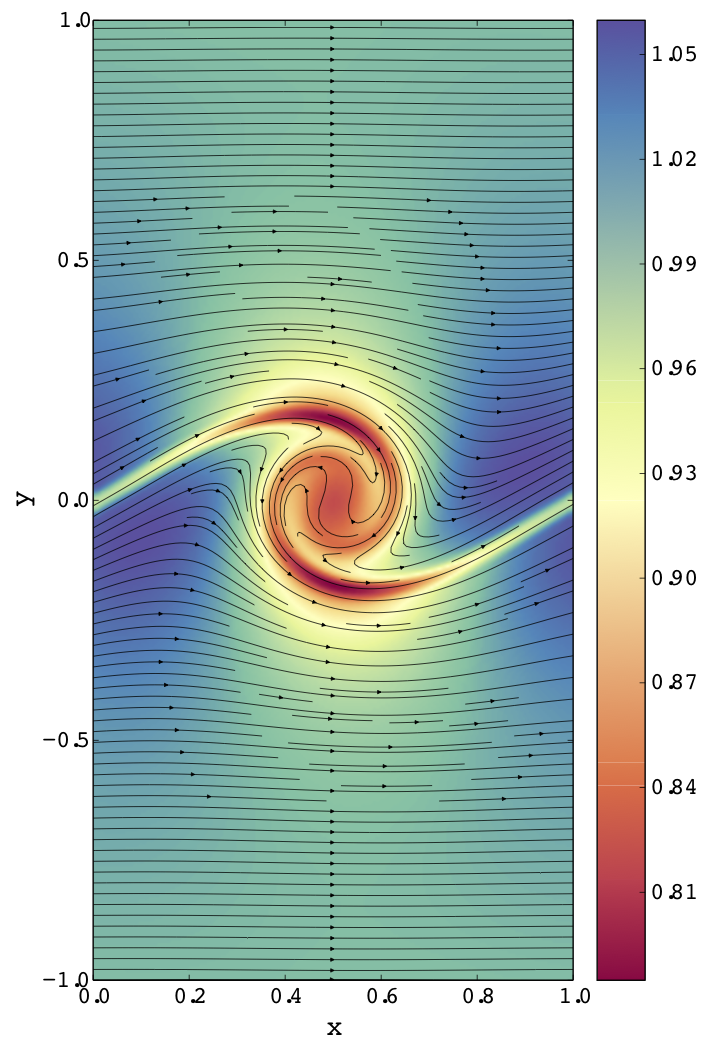} & 
\includegraphics[height=2.9in]{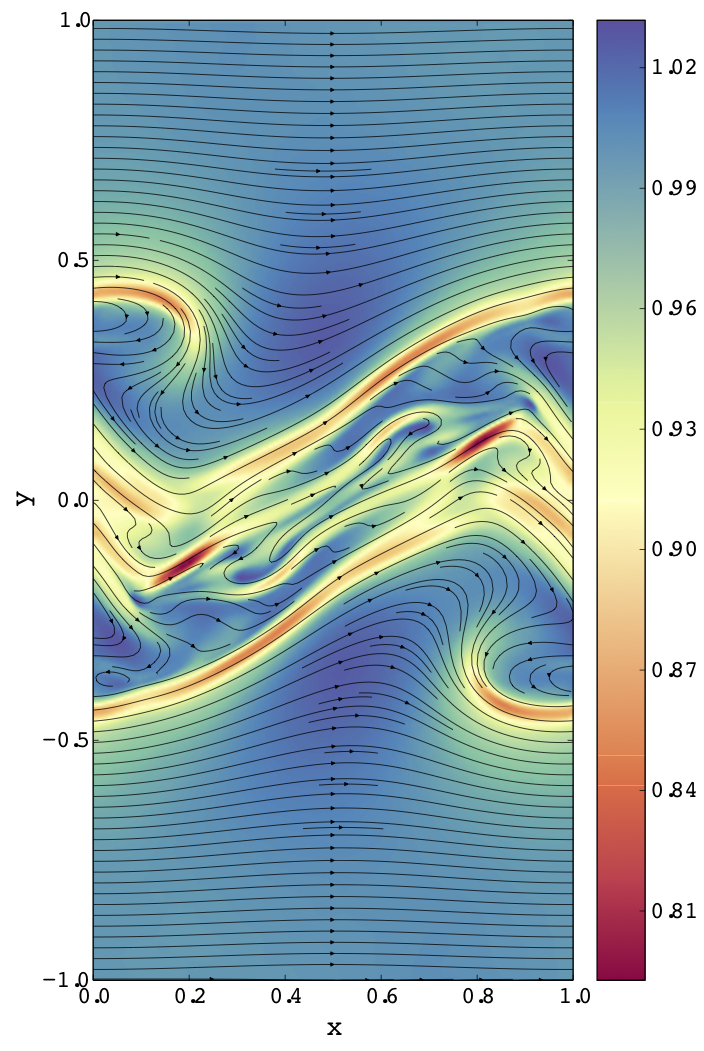} &
\includegraphics[height=2.9in]{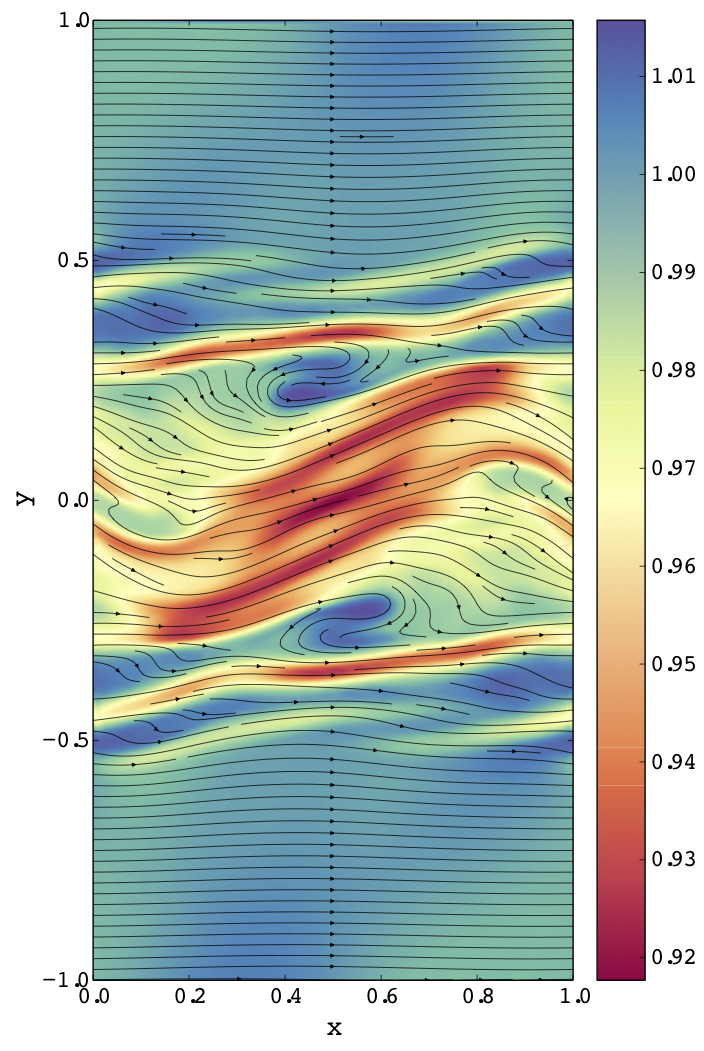}\\
\includegraphics[height=2.9in]{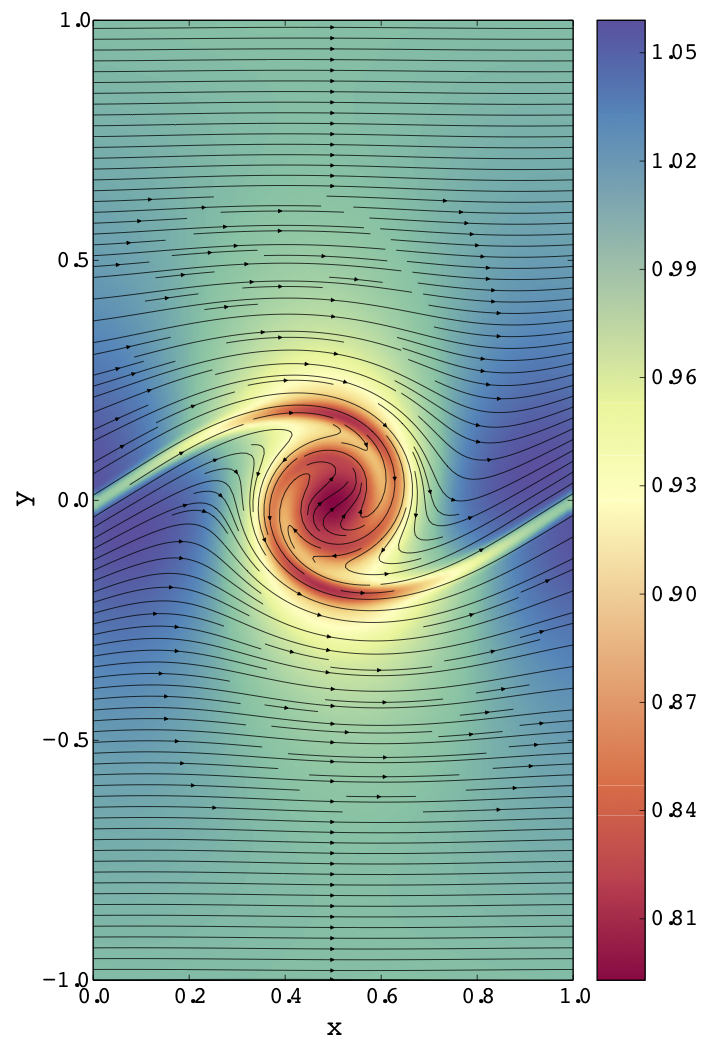} & 
\includegraphics[height=2.9in]{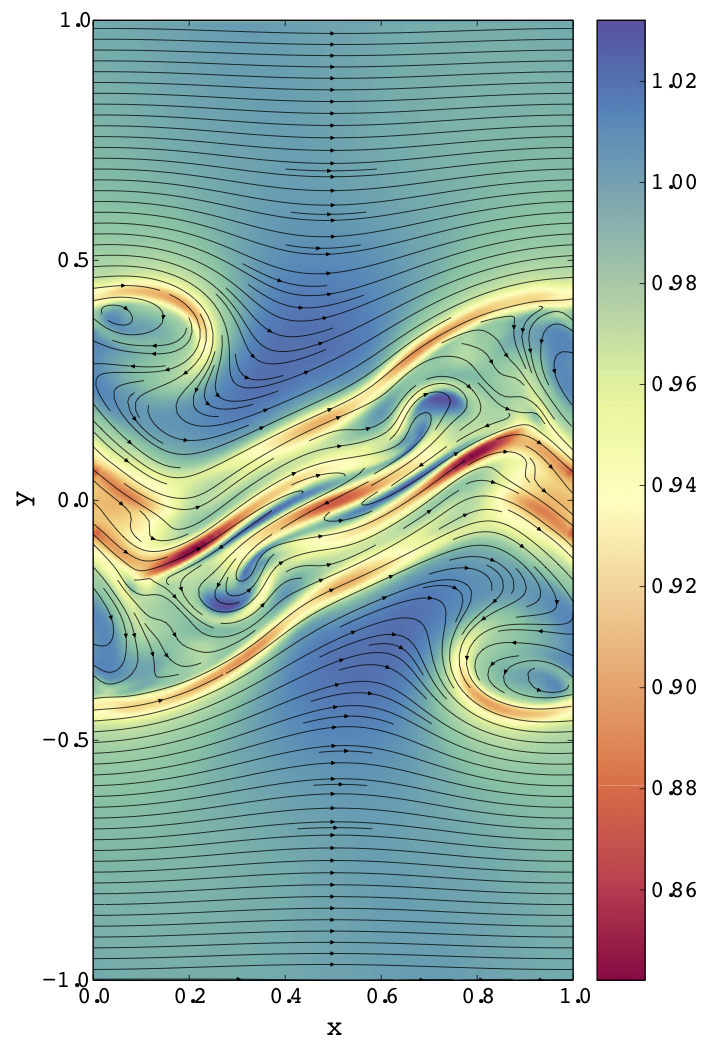} &
\includegraphics[height=2.9in]{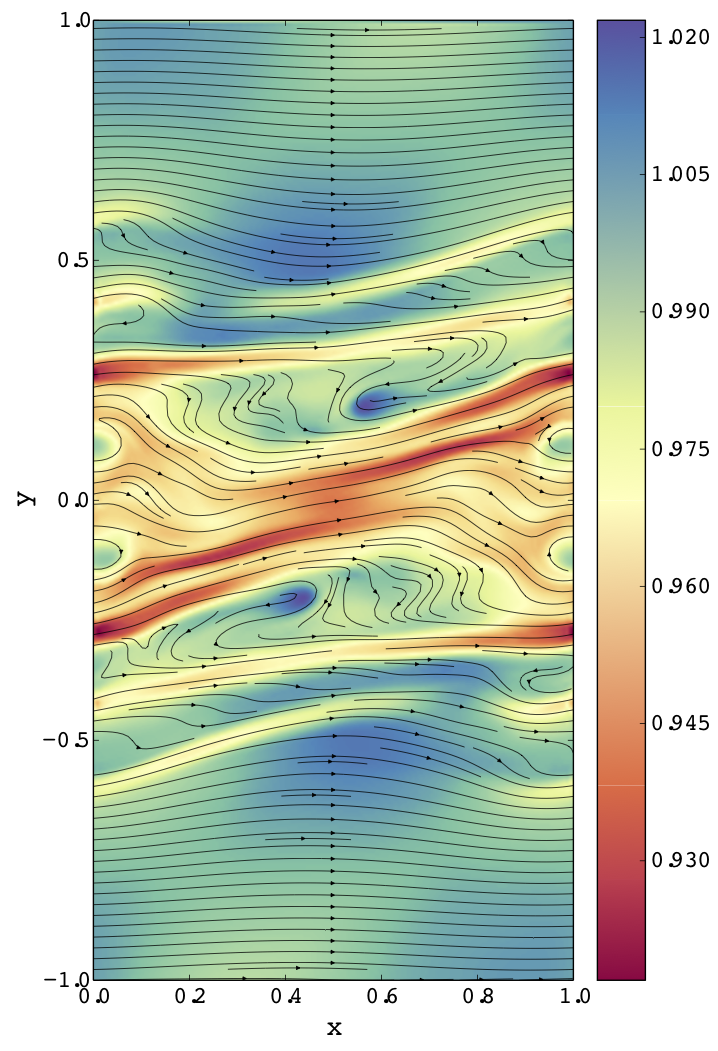}
\end{tabular}
\caption{Evolution of the rest mass density (colors) and the magnetic field lines for $\beta_x=0.02$. The left column are snapshots of the instability at $t=5$, the middle column at $t=10$, and the right column at $t=20$, where the system has reached a quasi stationary state. The top row is the evolution of a diamagnetic fluid with $\chi_m=-0.2$, the middle row is the usual MHD case without magnetic susceptibility, and the bottom row is the paramagnetic evolution with $\chi_m=0.2$.}
\label{rho3d}
\end{figure*}

In figure (\ref{rho3d}) we present the evolution of the rest-mass density and the magnetic field lines for three different simulations with $\beta_x=0.02$. The upper row corresponds to the diamagnetic case with $\chi_m=-0.2$ and the bottom one to the paramagnetic case with $\chi_m=0.2$. For comparison purposes, we show in the middle row the snapshots of the evolution with $\chi_m=0$. In the right column we present the instability at the end of the linear phase, when $t=5$. At this time there is a decrease at the center of the vortices that depends on the magnetic susceptibility. We notice that the vortex in the paramagnetic fluid has the lowest density of the three cases. Moreover, during the evolution (middle column with $t=10$) and at the end of it the morphology is very similar, but the density and magnetic field distributions change in the diamagnetic and paramagnetic case in comparison with the $\chi_m=0$ evolution. 

\begin{figure}
\begin{tabular}{l}
\includegraphics[height=1.15in]{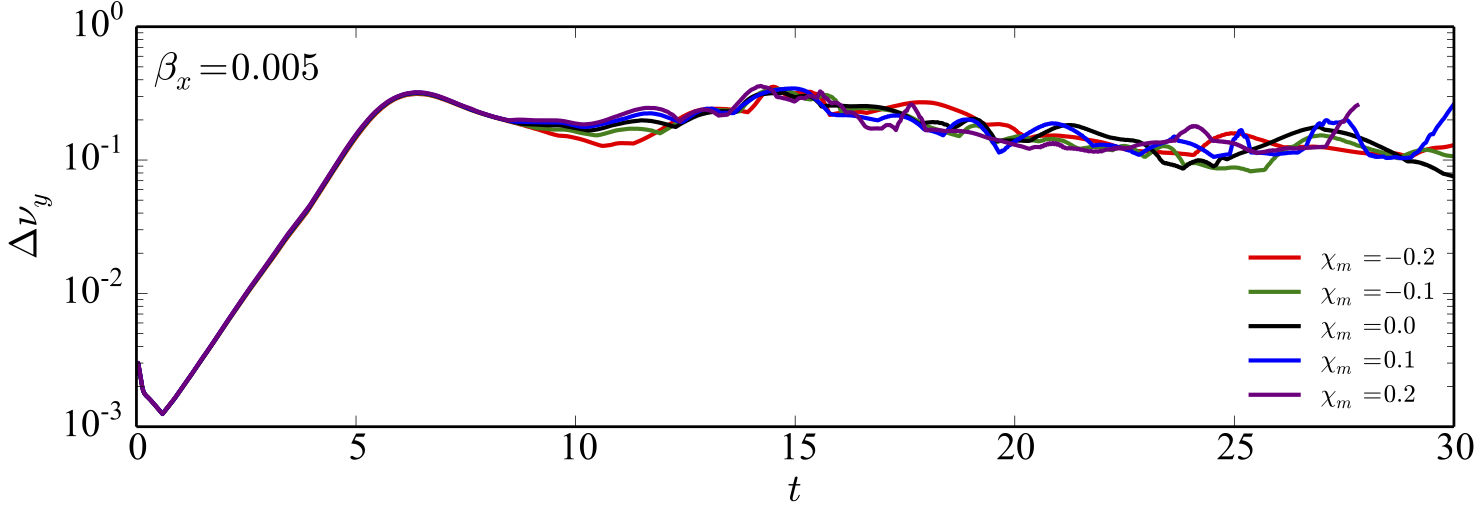}\\
\includegraphics[height=1.15in]{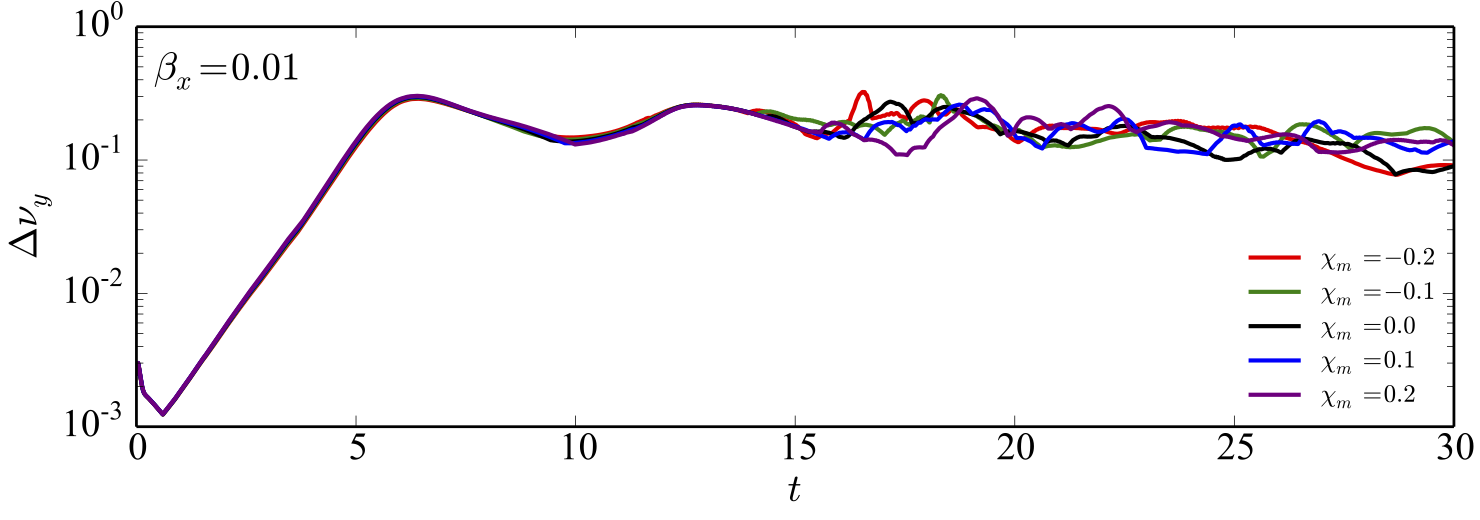}\\
\includegraphics[height=1.15in]{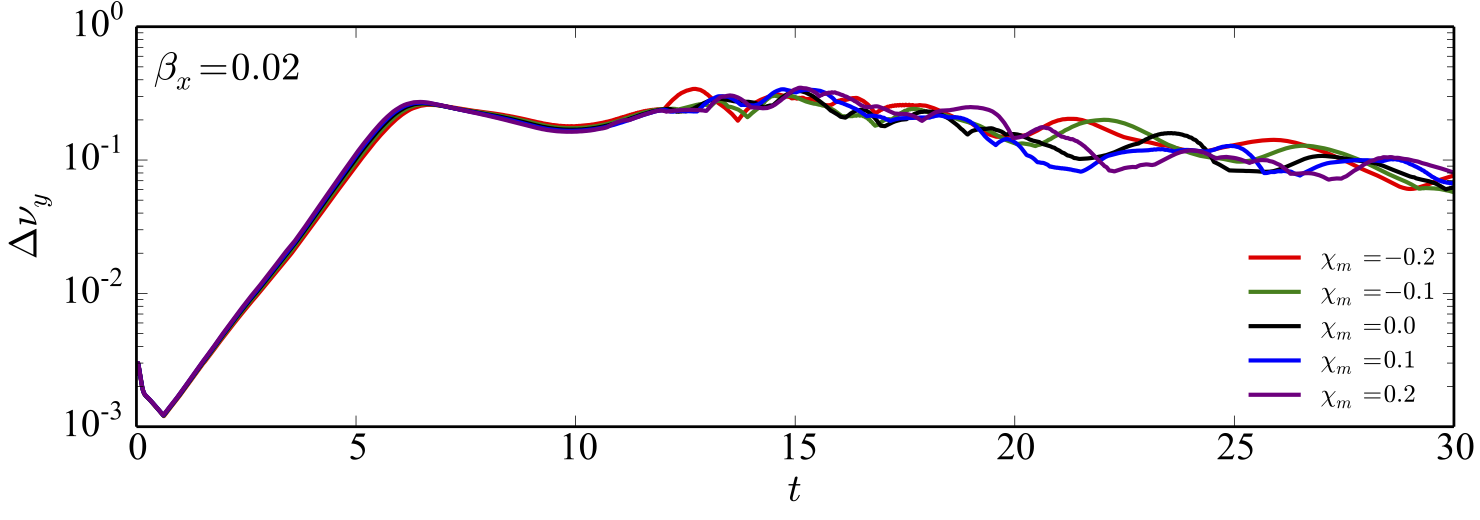}\\
\includegraphics[height=1.15in]{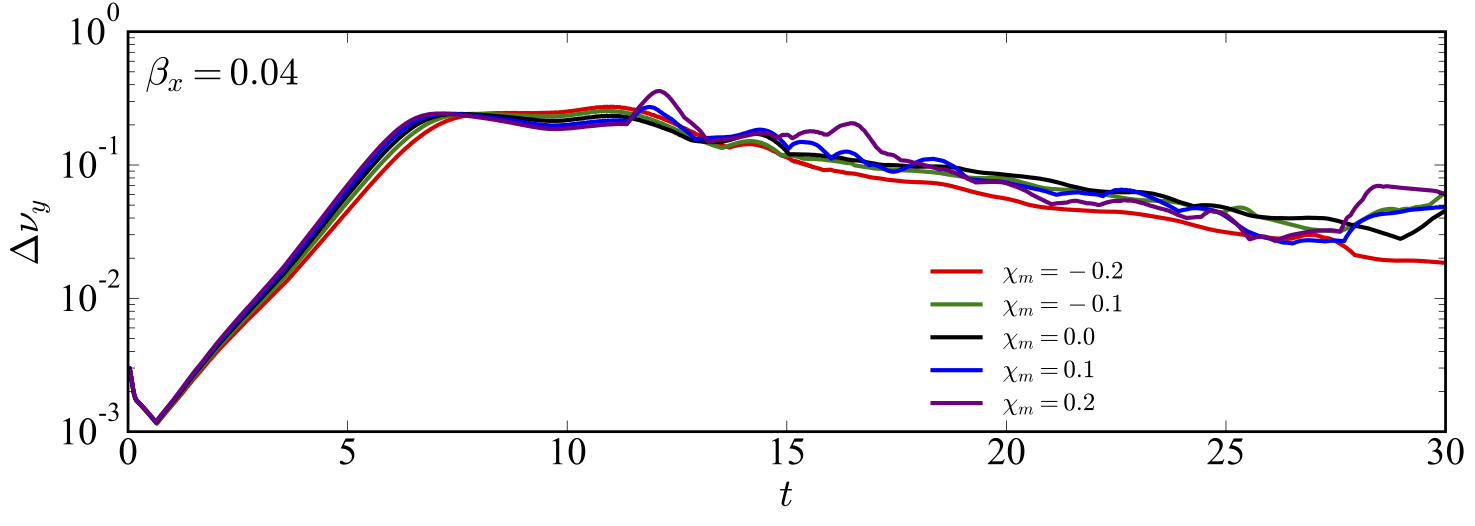}\\
\includegraphics[height=1.15in]{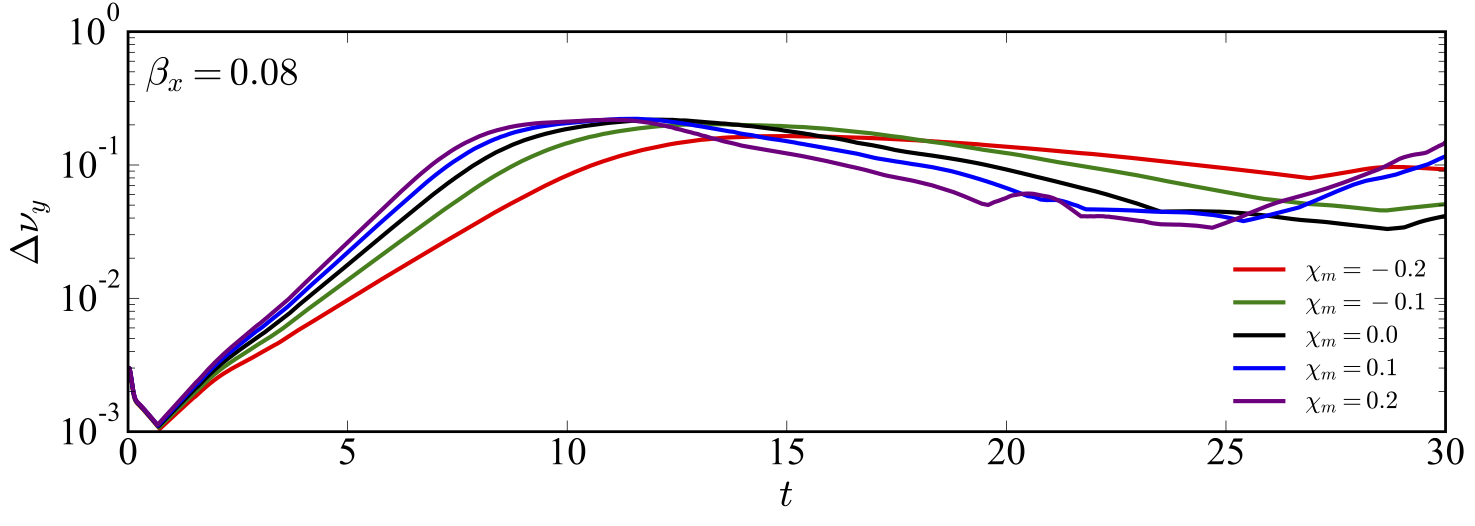}
\end{tabular}
\caption{Amplitude of the perturbation $\Delta v_{y}$ as a function of time for three magnetization parameters $\beta_x$. In each plot we present the results for two diamagnetic fluids with $\chi_m=-0.1, -0.2$, two paramagnetic fluids with $\chi_m=0.1, 0.2$, and for a fluid without magnetic susceptibility ($\chi_m=0$).}
\label{perturbation}
\end{figure}

A more detailed analysis of the development of the Kelvin-Helmholtz instability can be obtained by tracking the evolution of the the perturbation as it is done in \cite{1996ApJ...456..708M} or in \cite{2006A&A...454..393B}. In figure (\ref{perturbation}) we plot the amplitude of the perturbation, $\Delta v_y=0.5(v_{y_{\text{max}}}-v_{y_{\text{min}}})$, as a function of time for different values of the magnetization parameter, $\beta_x$. In each plot we vary the magnetic susceptibility of the fluids, such that we evolve for each $\beta_x$, two diamagnetic fluis, two paramagnetic fluids, and the case with $\chi_m=0$. The most interesting temporal region in each plot corresponds to the time interval $(0,5)$ since it represents the lineal evolution, and therefore we can compare our analytical results with the simulations.

First of all, it is important to mention that our simulations reproduce the slopes of the usual RMHD case without magnetic susceptibility \citep{1980MNRAS.193..469F, 2008A&A...490..493O}, so we are going to focus our analysis in the changes resulting from the inclusion of $\chi_m$. Now, as it was pointed out by \cite{2006A&A...454..393B}, the cases with $\beta=0.005$ and $\beta=0.01$ present an hydrodynamic behaviour in the linear evolution because the magnetic pressure is small enough in comparison with the gas pressure. A consequence of this fact is that the magnetic susceptibility does not contribute significantly in the growth rate, which is determined by the slope of the plot in the linear phase. This result was predicted by our linear analysis with the red and green curves in figure (\ref{growth_chi}). Nevertheless, from $\beta_x=0.02$, we notice that the slopes for the different $\chi_m$ values start to separate. The case with $\chi_m=0.08$ shows clearly the differences between the diamagnetic and paramagnetic evolutions in the linear regime. When we consider a diamagnetic fluid, the growth rate of the instability (slope of the curve) is remarkably reduced in comparison with the case without magnetic susceptibility. Therefore, based on the linear analysis and on the simulations, we can conclude that the diamagnetism stabilizes the interface between the fluids, while the paramagnetism destabilizes it. 

On the other hand, the plots for $\beta_x=0.04$ and $\beta_x=0.08$ in figure (\ref{perturbation}) show that the effect of the magnetic susceptibility on the growth rates (slopes) is greater when $\chi_m<0$. This sort of asymmetry between the paramagnetism and diamagnetism was also predicted by the analytical results, and is quite evident in the curve for $\beta_x=0.08$ of figure (\ref{growth_chi}) because the slope of that plot is steeper in $\chi_m<0$ than in the region $\chi_m>0$. As a final remark of figure (\ref{perturbation}), our simulations suggest that the global behaviour of $\Delta v_y$ in the non linear regime is statistically independent on the magnetic susceptibility.

\begin{figure}
\begin{tabular}{l}
\includegraphics[height=1.15in]{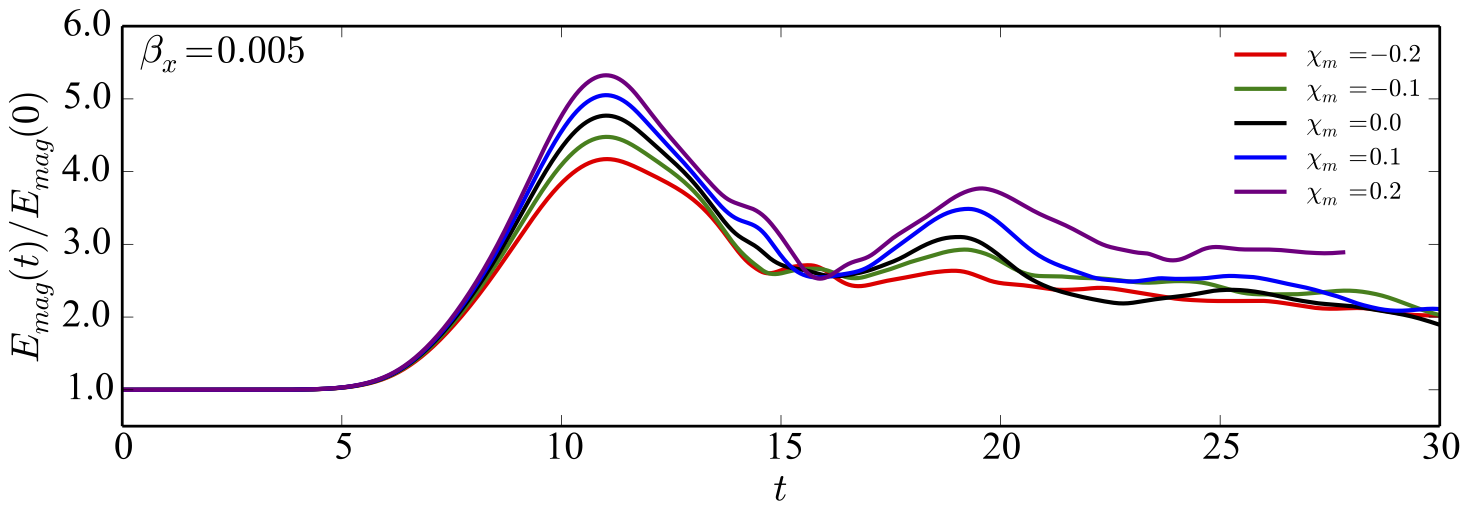}\\
\includegraphics[height=1.15in]{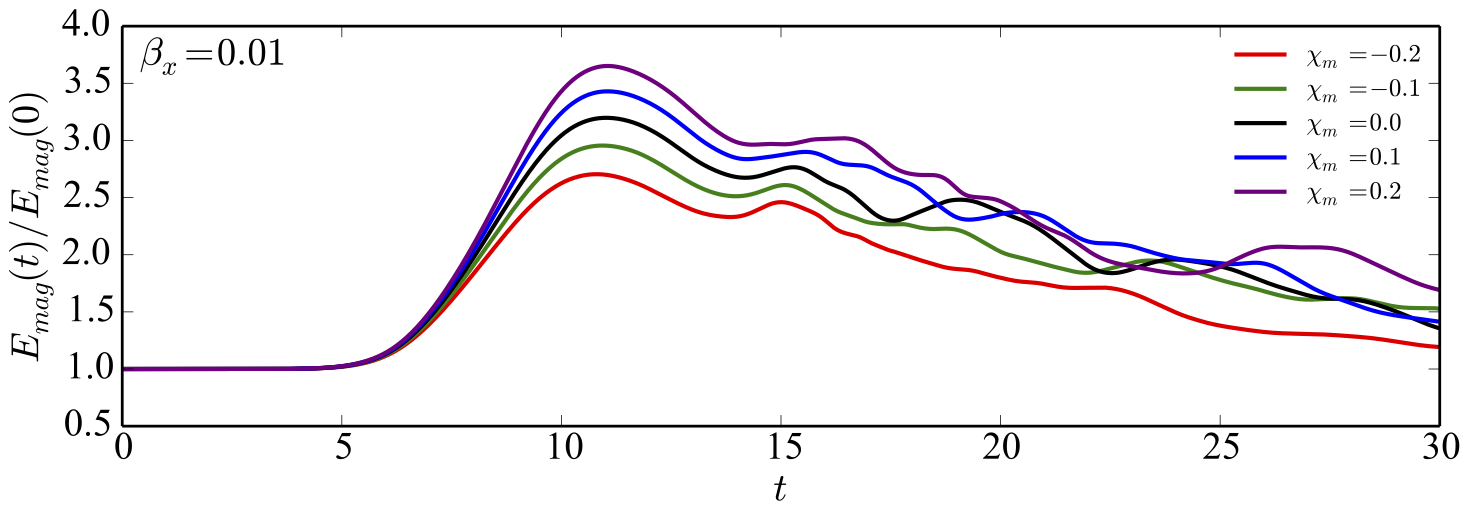}\\
\includegraphics[height=1.15in]{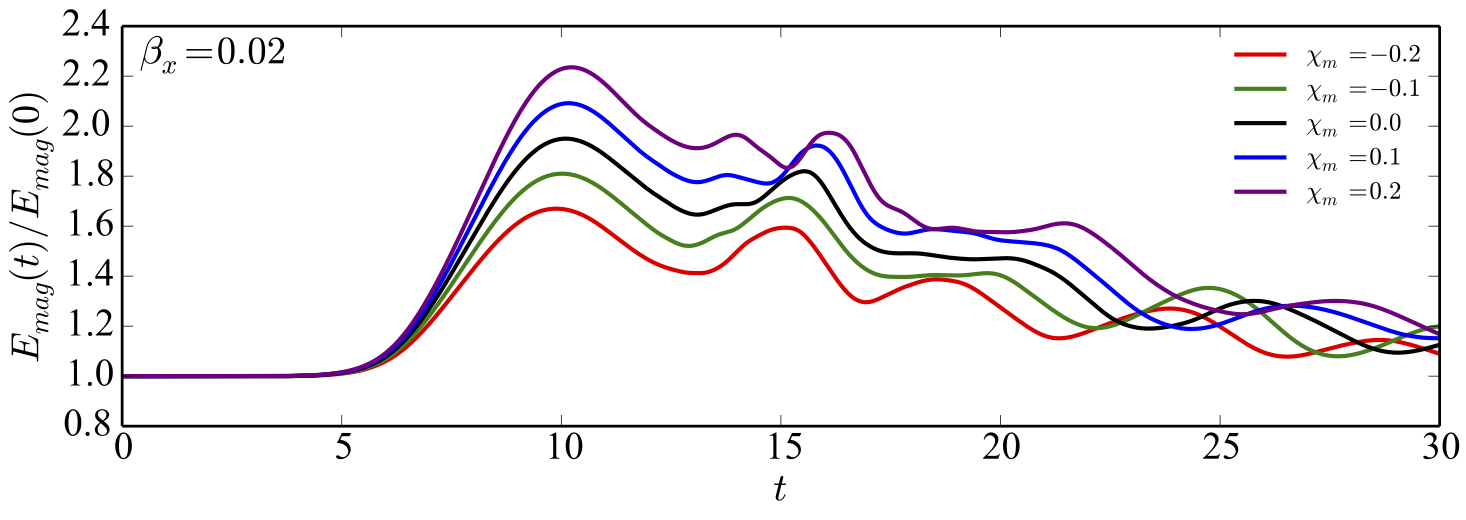}\\
\includegraphics[height=1.15in]{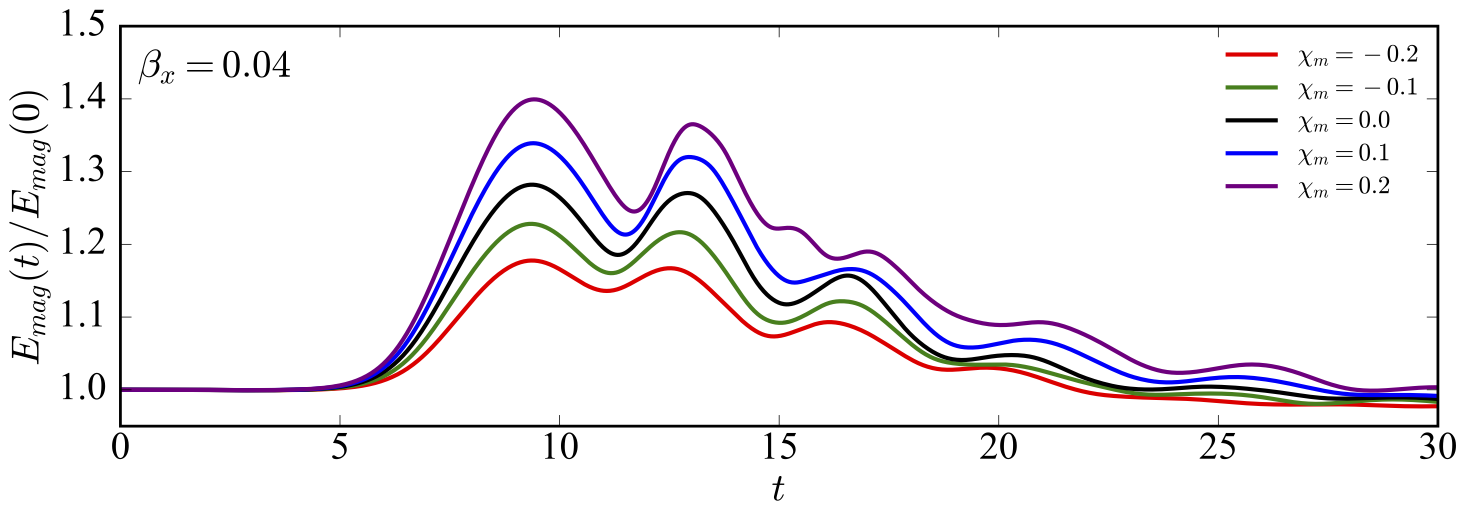}\\
\includegraphics[height=1.15in]{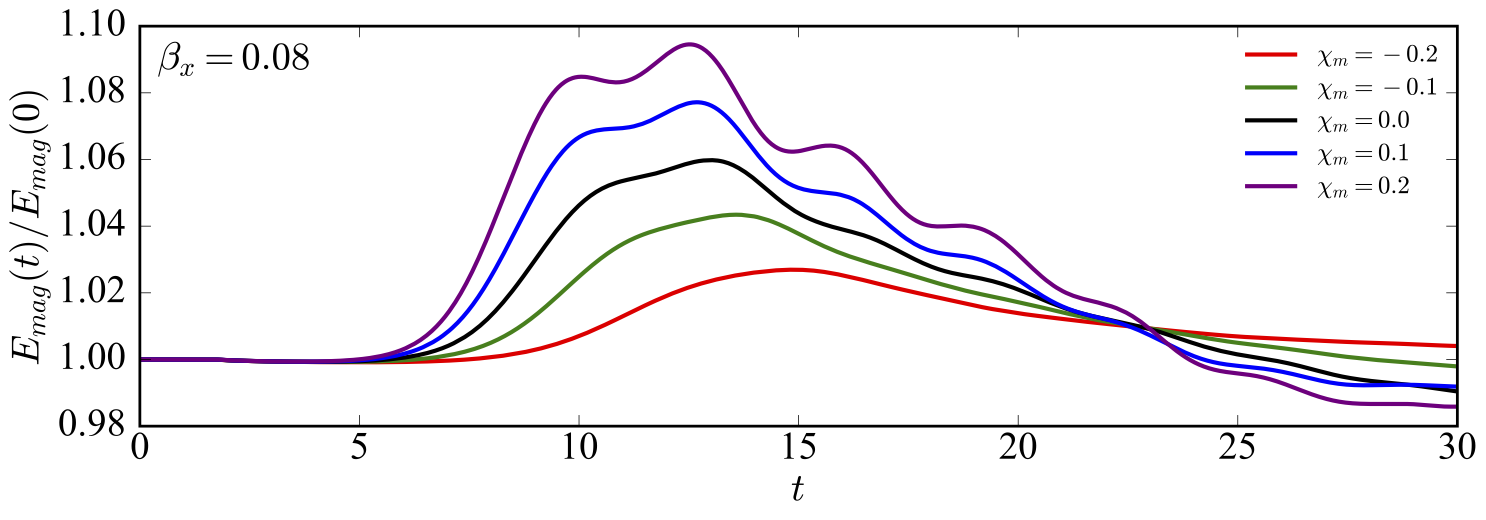}
\end{tabular}
\caption{Magnetic energy amplification $E_{mag}(t)/E_{mag}(0)$ as a function of time for three different values of magnetization parameter $\beta_x$. Each panel shows the results for different magnetic susceptibilities $\chi_m$.}
\label{ampli}
\end{figure}

Another interesting phenomena is the magnetic energy amplification \citep{1984JGR....89..801M}. This process is described by \cite{1996ApJ...456..708M}, and it is based on the fact that the magnetic field lines are attached to a high-conducting fluid, so the vortical motions stretch the lines, while the magnetic field is concentrated and amplified inside the narrow filaments. The amplification ends when the gradient scales of the field becomes of the order of the grid resolution, so numerical resistivity acts by producing magnetic reconnection events. Figure (\ref{ampli}) is the same as figure (\ref{perturbation}), but this time for the magnetic energy amplification $E_{\text{mag}}(t)/E_{\text{mag}}(0)$, where
\begin{equation}
E_{\text{mag}}(t)=\frac{1}{2}\int B(t)^2dA,
\label{amplification}
\end{equation} 
is the total magnetic energy in our 2D numerical space. In this figure, we see that $E_{\text{mag}}$ is more amplified when the instability takes place between paramagnetic fluids. This behaviour is very interesting because even in the approximatelly hydrodynamic case where $\beta=0.005$, the magnetic energy amplification is affected by the magnetic susceptibility in the non linear regime. The maximum amplification is obtained in the non linear phase and coincides with the first reconnection in the cases with $\beta_x\leq 0.04$, but in the last case (bottom panel), the maximum amplification seems to be obtained in the second reconnection event.

\begin{figure}
\begin{tabular}{c}
\includegraphics[height=2.5in]{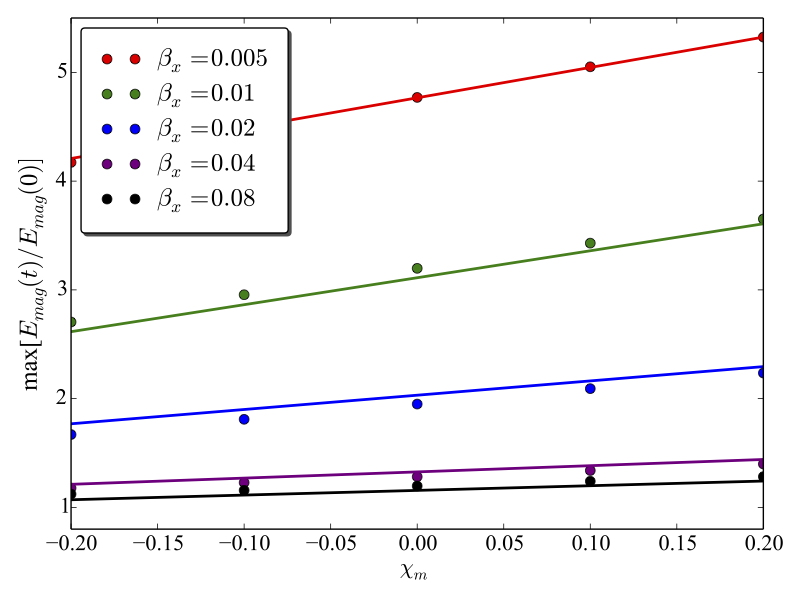}
\end{tabular}
\caption{Maximum amplification of the magnetic energy as a function of $\chi_m$ for different values of $\beta_x$. The lines correspond to the best fit, according to equation (\ref{best_fit}).}
\label{ampli_fix}
\end{figure}

The maximum of the magnetic energy amplification for each $\beta_x$ was plotted as a function of the magnetic susceptibility in figure (\ref{ampli_fix}). The dots are the values obtained from the simulations and the lines are the best fit computed through the expression,
\begin{equation}
\text{Max}\left[\frac{E_{\text{mag}}(t)}{E_{\text{mag}}(0)}\right]=\frac{m_1\beta_x^{~2}\chi_m}{(m_2^{~2}+\beta_x^{~2})^{\kappa_1}}+\frac{m_3}{\beta_x^{~\kappa_2}},
\label{best_fit}
\end{equation}
where $m_1=0.0085469$, $m_2=0.00506871$, $m_3=0.183125$, $\kappa_1=1.65678$, and $\kappa_2=0.615124$. The last equation describes the maximum amplification of the magnetic energy in terms of the magnetization parameter at $t=0$ and the magnetic susceptibility of the fluids. In figure (\ref{ampli_fix}) we notice that the slopes of the curves increase when $\beta_x$ is reduced, so the magnetic susceptibility is more important for small values of $\beta_x$. If some sort of dynamo process would operate in the system, the magnetic susceptibility of the fluids could make the RKHI a more efficient and effective magnetic field seed amplification mechanism \citep{2009ApJ...692L..40Z}, which could be described by the growth rates of section \ref{sec3} along with the equation (\ref{best_fit}).

Finally, we plot the integrated velocity across the shear layer as a function of $y$ for different values of $\beta_x$ at $t=25$. At this time the system has reached a stationary state and the mixing layer has completely formed. In each pannel we use the same values for $\chi_m$ as in the previous figures (\ref{perturbation}, \ref{ampli}). As we can observe, the shear layer width is rougly independent of the magnetic susceptibility.

\begin{figure*}
\begin{center}
\begin{tabular}{cc}
\includegraphics[height=2.05in]{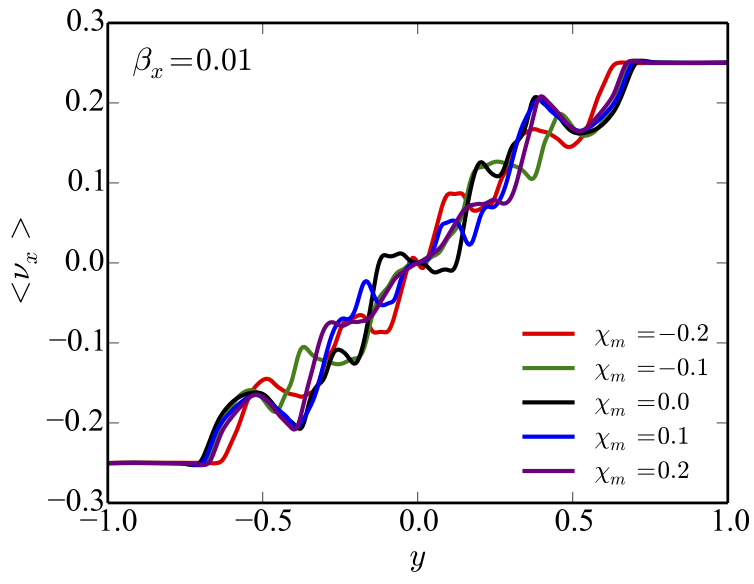} & 
\includegraphics[height=2.05in]{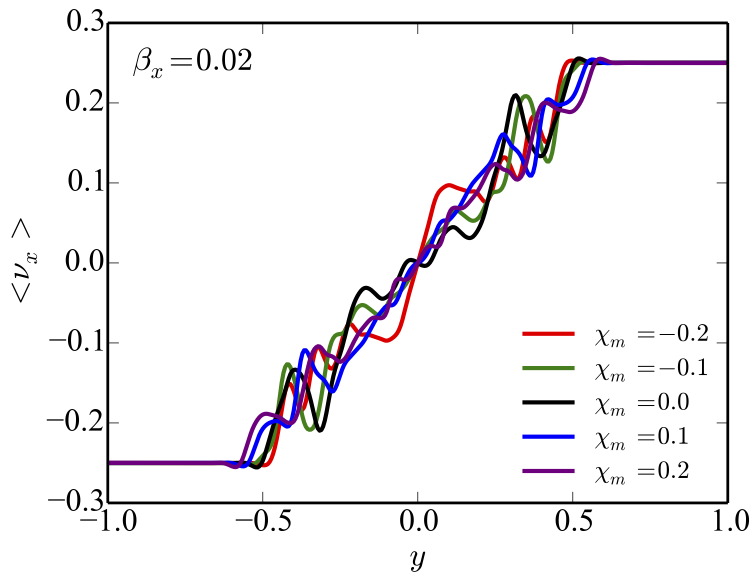}\\
\includegraphics[height=2.05in]{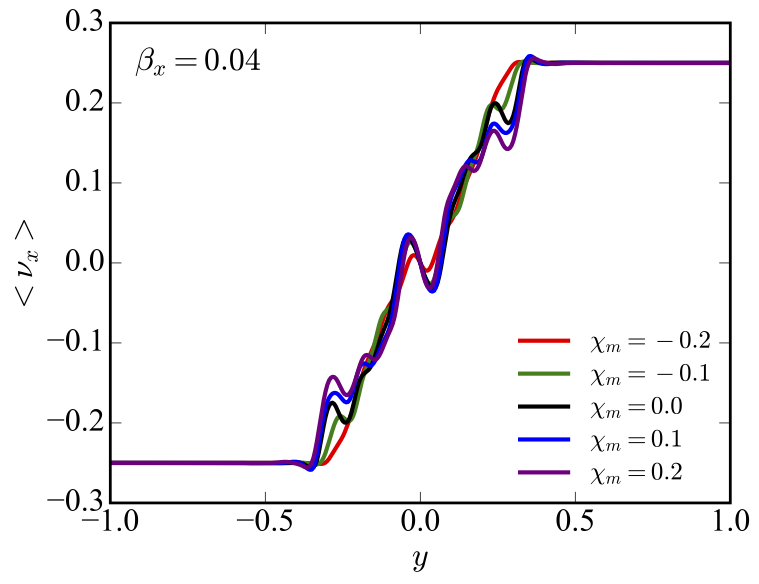} &
\includegraphics[height=2.05in]{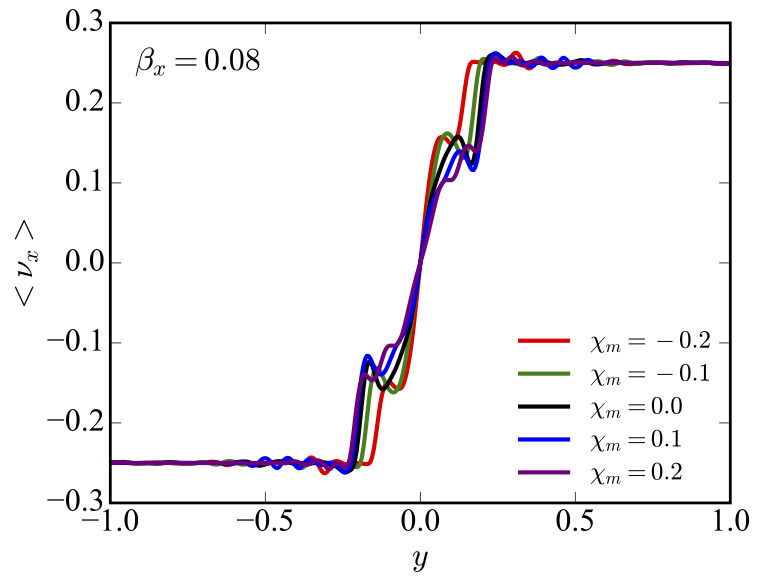}
\end{tabular}
\caption{Integrated velocity profile at $t=25$ as a function of $y$ for different values of $\beta_x$ and $\chi_m$. At this time the system has reached a quasi-steady state.}
\label{beta}
\end{center}
\end{figure*}




\section{Conclusions}
\label{sec5}

In this paper we have explored the effect of the magnetic susceptibility in the relativistic KHI between two uniform and magnetized fluids moving with opposite velocities. The fluids are separated by a planar interface located at $y=0$ and the magnetic field is assumed to be uniform in all the space and parallel to the flow velocity. The linear analysis was carried out for a perturbation that propagates in an arbitrary direction, but the results were obtained for a wave vector  parallel to the $x-y$ plane. We first considered the linear evolution of the instability, so we compute the analytical dispersion relation and found that the Alfv$\grave{\text{e}}$n modes do not contribute to the instability of the interface. The only unstable modes are associated with magneto-acustic waves.

By solving the dispersion relation we found that the planar interface between diamagnetic fluids is more stable than in the original case without magnetic susceptibility. On the contrary, when the paramagnetism is present in the fluids, the Kelvin-Helmholtz instability grows faster than in the previous cases ($\chi_m=0$ and $\chi_m<0$). This effect is higher when the magnetic pressure dominates over the gas pressure. Additionally, we found that the interval of possible values for the relativistic Mach number increases with the magnetic susceptibility. In this respect, the most significant differences in the growth rates between the diamagnetic and the paramagnetic cases are found in the stability threshold.

Subsequently, we use CAFE code to test the analytical results mentioned in the last paragraph, and obtain the non linear evolution of the instability. In this regime the magnetic field is amplified, so the kinetic energy associated with the shear flow is transformed into magnetic energy. We found, by simulating fluids with different magnetic susceptibilities, that the magnetic energy is amplified to higher values in paramagnetic fluids than in diamagnetic ones. Surprisingly, we found that the effect of the magnetic susceptibility in the field amplification is larger in the cases with small $\beta_x$. It suggests that the amplification of small seed magnetic fields in the KH instability could be more effective and efficient if the flow presents paramagnetic properties. Therefore, we obtain form the numerical simulations a potentially useful expression for the magnetic energy amplification in terms of the initial $\beta_x$ and $\chi_m$. Finally, we computed the integrated velocity profile in the same direction of the flows and showed that the shear layer width is almost independent of the magnetic susceptibility of the fluids.

\section*{Acknowledgements}

We highly appreciate the fruitful discussion and suggestions from the anonymous referee, which have improved this work. O. M. P. wants to thanks the financial support from COLCIENCIAS under the program Becas Doctorados Nacionales 647 and Universidad Industrial de Santander. F.D.L-C was supported in part by VIE-UIS, under Grant No. 2493 and by COLCIENCIAS, Colombia, under Grant No. 8863.





\appendix

\section{Differences between the RMHD with a renormalized magnetic field and the RMHD with magnetic susceptibility}
\label{apenA}
According to the results of sections (\ref{sec3}) and (\ref{sec4}), it seems that the main effect of $\chi_m$ is to increase or reduce the intensity of the magnetic field in the system. In fact, if we look at the energy-momentum tensor (\ref{energy_tensor}), we may try to renormalize the magnetic field as
\begin{equation}
b^{\mu}\rightarrow \sqrt{1-\chi}b^{\mu},
\label{renormalization_opt}
\end{equation}
in order to include $\chi_m$ into the new field and be able to use the existing RMHD codes and well-known linear analysis to describe the magnetic susceptibility effects. Nevertheless, as it was mentioned in section (\ref{sec2}), such a renormalization cannot be done because of the coefficient $1-2\chi$ in the pressure $p^{*}$. This coefficient is important for the magneto-acustic modes obtained from (\ref{magnetoacustic_equation}) which are indeed responsible for the KHI. In this appendix we discuss the differences between the usual RMHD with a renormalized magnetic field, and the RMHD with magnetic susceptibility including the complete form of the energy-momentum tensor (\ref{energy_tensor}).

\begin{figure}
\begin{tabular}{c}
\includegraphics[height=2.5in]{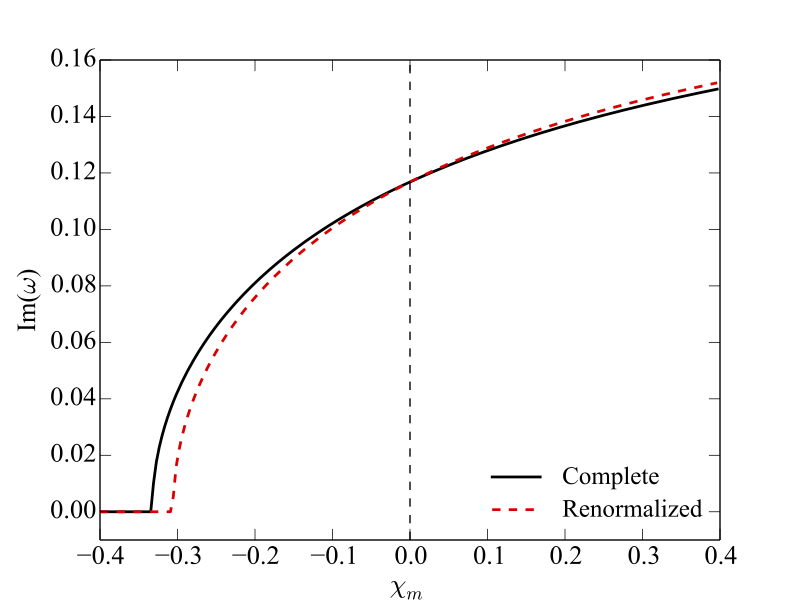}
\end{tabular}
\caption{Growth rate comparison between the purely RMHD with a renormalized magnetic field ($b^{\mu}\rightarrow \sqrt{1-\chi}b^{\mu}$) and the complete RMHD with magnetic susceptibility for a magnetization parameter $\beta_x=0.1$. The most significant differences are obtained for diamagnetic fluids in the stability threshold.}
\label{apen_growth}
\end{figure}

From the linear theory described in section (\ref{sec3}) we compute the growth rate as a function of the magnetic susceptibility for a magnetization parameter $\beta_x=0.1$. The result of this exercise, presented as the black curve in figure (\ref{apen_growth}), describes the complete effect of the magnetic polarization in the KHI since we use $T^{\mu\nu}$ as given in (\ref{energy_tensor}). We note that for $\chi_m\lesssim -0.331$, the KHI is suppresed by an increase of the magnetic pressure due to the polarization of the fluid. Now, doing $\chi_m=0$ in all the expressions of section (\ref{sec3}) we recover the usual RMHD analysis. If in addition we try to emulate the effect of the magnetic susceptibility though the renormalization (\ref{renormalization_opt}), then the growth rate for the $\beta_x=0.1$ case behaves like the red-dashed line of figure (\ref{apen_growth}). In this case the KHI is suppresed when $\chi_m\lesssim -0.305$, so there is an interval of $\chi_m$ for wich the RMHD with a renormalized magnetic field predicts stability while the RMHD with magnetic susceptibility (black line) predicts instability. In fact, the higher differences between the two cases occur in the stability threshold.

\begin{figure}
\begin{tabular}{c}
\includegraphics[height=2.5in]{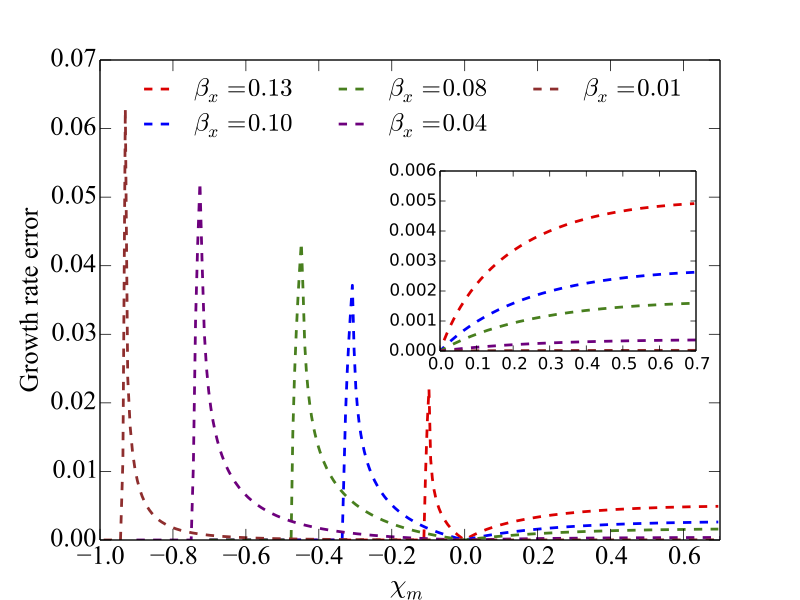}
\end{tabular}
\caption{Absolute error of the growth rate obtained between the RMHD with magnetic susceptibility and the purely RMHD with a renormalized magnetic field for different magnetization parameters. When $\chi_m<0$ the error in the istability threshold is larger when $\beta_x$ is small, and when $\chi_m>0$ the error grows with $\beta_x$.}
\label{apen_error}
\end{figure}

The absolute error of the growth rate, taking into account that the RMHD with magnetic susceptibility predicts the true values of Im$(\omega)$, is presented in figure (\ref{apen_error}) for different magnetization parameters. In the case of diamagnetic fluids, the error grows when $\beta_x$ is reduced, but the biggest difference between the two cases can be noted for a small enough $\chi_m$. On the contrary, in a system with a sufficiently large magnetization parameter, the error can be observed with a low $\chi_m$. In the paramagnetic fluids the error increases with $\beta_x$ but it is in general smaller than in the diamagnetic cases. 
\begin{figure}
\begin{tabular}{l}
\includegraphics[height=1.15in]{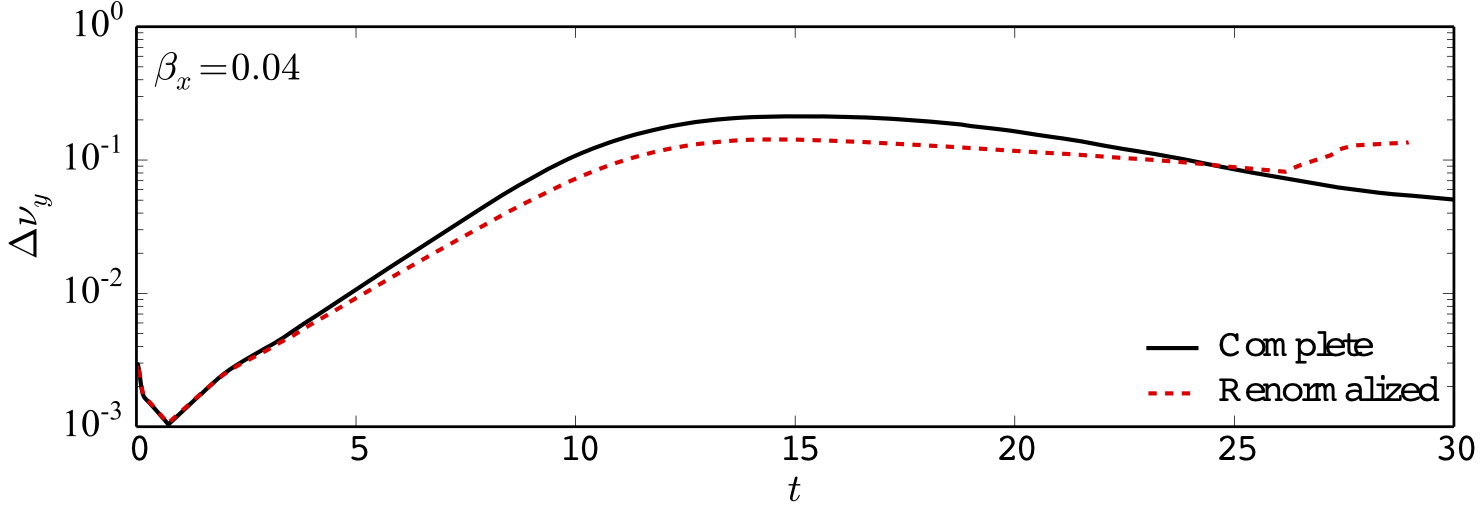}\\
\includegraphics[height=1.15in]{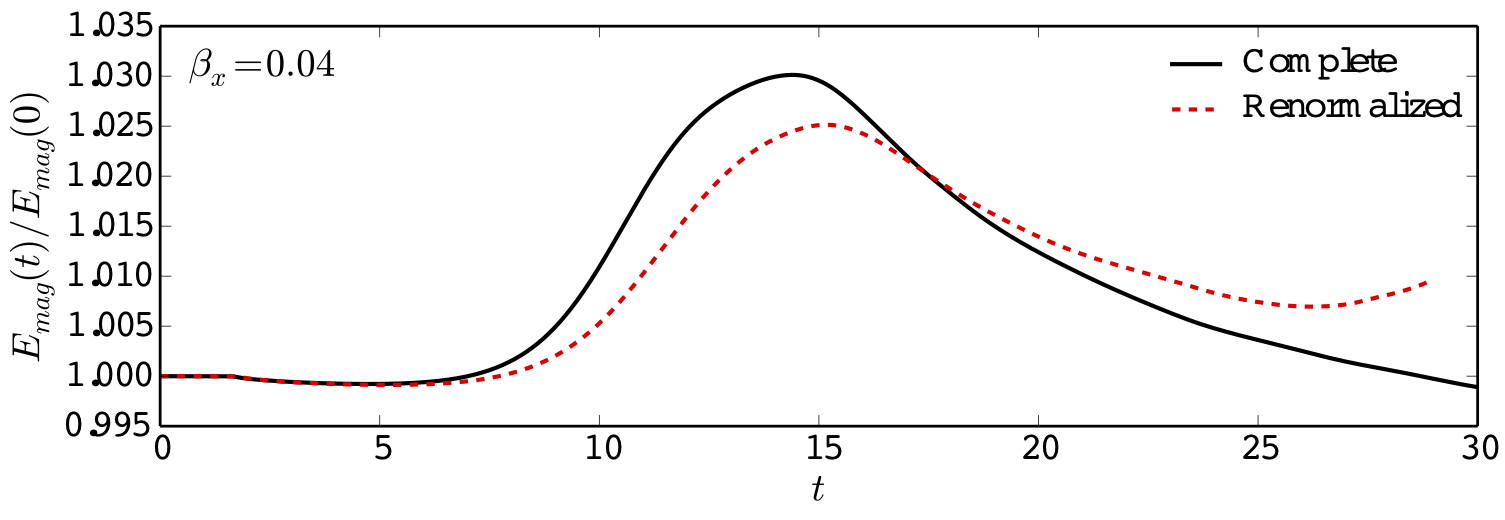}
\end{tabular}
\caption{Amplitude of the perturbation, $\Delta v_y$, (top) and magnetic energy amplification, $E_{mag}(t)/E_{mag}(0)$, (bottom) as a function of time for $\beta_x=0.04$ and $\chi_m=-0.6$. The black curves correspond to the simulation with the RMHD with magnetic susceptibilty that we present in this work, and the red-dashed lines correspond to the purely RMHD with a renormalized magnetic field.}
\label{response}
\end{figure}

In addition to the linear analysis, we simulate the evolution of the KHI with $\beta_x=0.04$ and $\chi_m=-0.6$, so the error in the growth rate is not that big as in the instability thereshold but it is comparable with the expected error in the paramagnetic cases (refer to figure \ref{apen_error}). In figure (\ref{response}) we plot the amplitude of the perturbation and the magnetic energy amplification as a function of time. The black curve again correspond to the RMHD with magnetic polarization that we present in this work, and the red-dashed line correspond to the purely RMHD with a renormalized magnetic field. We note that, as predicted by the linear analysis, the instability grows faster when we solve the RMHD equations with the energy-momentum tensor given in (\ref{energy_tensor}) than solving the purely RMHD equations with the renormalized magnetic field (\ref{renormalization_opt}).

Now, despite the fact that the instability seems to saturate approximately at the same time in both cases, the maximum amplification of the magnetic energy is reached at a later time in the renormalized case, and its value is lower than the complete RMHD with magnetic polarization. The relative error between both cases is $\approx 0.48\%$; nevertheless, if we reduce the magnetization parameter to $0.01$, the relative error increases considerably to $\approx 3.36\%$ for a paramagnetic fluid with $\chi_m=0.4$. Therefore, there are some cases in which the purely RMHD with a renormalized magnetic field does not lead to the correct results. In particular, the growth rates, the maximum magnetic energy amplification, and the time at which this maximum is reached.


\bsp	\label{lastpage}
\end{document}